\documentclass[iop]{emulateapj}

\newcommand\papercls{y} 

\usepackage{apjfonts}
\usepackage{ifthen}
\usepackage{natbib}
\usepackage{amssymb, amsmath}
\usepackage{appendix}
\usepackage{etoolbox}
\usepackage{color}
\usepackage{makecell}
\if\papercls y
\newcommand\aas{\ref@jnl{AAS Meeting Abstracts}}
\newcommand\dps{\ref@jnl{AAS/DPS Meeting Abstracts}}
\newcommand\maps{\ref@jnl{MAPS}}
\else
\usepackage{astjnlabbrev-jh}
\fi

\if\papercls y
\bibpunct[, ]{(}{)}{,}{a}{}{,}
\else
\bibpunct[, ]{(}{)}{,}{a}{}{,}
\fi

\usepackage[
bookmarks=true,           
bookmarksnumbered=true,   
colorlinks=true,          
citecolor=blue,           
linkcolor=blue,           
menucolor=blue,           
urlcolor=blue,            
linkbordercolor={0 0 1},  
pdfborder={0 0 1},
frenchlinks=true]{hyperref}

\providecommand{\adsurl}[1]{\href{#1}{ADS}}

\makeatletter
\patchcmd{\NAT@citex}
  {\@citea\NAT@hyper@{%
     \NAT@nmfmt{\NAT@nm}%
     \hyper@natlinkbreak{\NAT@aysep\NAT@spacechar}{\@citeb\@extra@b@citeb}%
     \NAT@date}}
  {\@citea\NAT@nmfmt{\NAT@nm}%
   \NAT@aysep\NAT@spacechar\NAT@hyper@{\NAT@date}}{}{}

\patchcmd{\NAT@citex}
  {\@citea\NAT@hyper@{%
     \NAT@nmfmt{\NAT@nm}%
     \hyper@natlinkbreak{\NAT@spacechar\NAT@@open\if*#1*\else#1\NAT@spacechar\fi}%
       {\@citeb\@extra@b@citeb}%
     \NAT@date}}
  {\@citea\NAT@nmfmt{\NAT@nm}%
   \NAT@spacechar\NAT@@open\if*#1*\else#1\NAT@spacechar\fi\NAT@hyper@{\NAT@date}}
  {}{}
\makeatother

\makeatletter
\DeclareRobustCommand{\lowcase}[1]{\@lowcase#1\@nil}
\def\@lowcase#1\@nil{\if\relax#1\relax\else\MakeLowercase{#1}\fi}
\makeatother

\DeclareSymbolFont{UPM}{U}{eur}{m}{n}
\DeclareMathSymbol{\umu}{0}{UPM}{"16}
\let\oldumu=\umu
\renewcommand\umu{\ifmmode\oldumu\else\math{\oldumu}\fi}

\newcommand\chisq{\ifmmode{\chi\sp{2}}\else\math{\chi\sp{2}}\fi}
\newcommand\redchisq{\ifmmode{ \chi\sp{2}\sb{\rm red}}
                    \else\math{\chi\sp{2}\sb{\rm red}}\fi}

\let\oldsim=\sim
\renewcommand\sim{\ifmmode\oldsim\else\math{\oldsim}\fi}
\let\oldpm=\pm
\renewcommand\pm{\ifmmode\oldpm\else\math{\oldpm}\fi}
\newcommand\by{\ifmmode\times\else\math{\times}\fi}

\newcommand\tablebox[1]{\begin{tabular}[t]{@{}l@{}}#1\end{tabular}}
\newbox{\wdbox}
\renewcommand\c{\setbox\wdbox=\hbox{,}\hspace{\wd\wdbox}}
\renewcommand\i{\setbox\wdbox=\hbox{i}\hspace{\wd\wdbox}}

\newcount\timect
\newcount\hourct
\newcount\minct
\newcommand\now{\timect=\time \divide\timect by 60
         \hourct=\timect \multiply\hourct by 60
         \minct=\time \advance\minct by -\hourct
         \number\timect:\ifnum \minct < 10 0\fi\number\minct}

\catcode`@=11

\newcommand\comment[1]{}

\newcommand\commenton{\catcode`\%=14}
\newcommand\commentoff{\catcode`\%=12}

\renewcommand\math[1]{$#1$}
\newcommand\mathshifton{\catcode`\$=3}
\newcommand\mathshiftoff{\catcode`\$=12}

\comment{alignment tab}
\let\atab=&
\newcommand\atabon{\catcode`\&=4}
\newcommand\ataboff{\catcode`\&=12}

\let\oldmsp=\sp
\let\oldmsb=\sb
\def\sp#1{\ifmmode
           \oldmsp{#1}%
         \else\strut\raise.85ex\hbox{\scriptsize #1}\fi}
\def\sb#1{\ifmmode
           \oldmsb{#1}%
         \else\strut\raise-.54ex\hbox{\scriptsize #1}\fi}
\newbox\@sp
\newbox\@sb
\def\sbp#1#2{\ifmmode%
           \oldmsb{#1}\oldmsp{#2}%
         \else
           \setbox\@sb=\hbox{\sb{#1}}%
           \setbox\@sp=\hbox{\sp{#2}}%
           \rlap{\copy\@sb}\copy\@sp
           \ifdim \wd\@sb >\wd\@sp
             \hskip -\wd\@sp \hskip \wd\@sb
           \fi
        \fi}
\def\msp#1{\ifmmode
           \oldmsp{#1}
         \else \math{\oldmsp{#1}}\fi}
\def\msb#1{\ifmmode
           \oldmsb{#1}
         \else \math{\oldmsb{#1}}\fi}
\def\supon{\catcode`\^=7}
\def\supoff{\catcode`\^=12}
\def\subon{\catcode`\_=8}
\def\suboff{\catcode`\_=12}
\def\supsubon{\supon \subon}
\def\supsuboff{\supoff \suboff}

\newcommand\actcharon{\catcode`\~=13}
\newcommand\actcharoff{\catcode`\~=12}

\newcommand\paramon{\catcode`\#=6}
\newcommand\paramoff{\catcode`\#=12}

\comment{And now to turn us totally on and off...}

\newcommand\reservedcharson{ \commenton  \mathshifton  \atabon  \supsubon 
                             \actcharon  \paramon}

\newcommand\reservedcharsoff{\commentoff \mathshiftoff \ataboff \supsuboff 
                             \actcharoff \paramoff}

\catcode`@=12
\reservedcharsoff

\comment{End of joe.tex character unreservations.}

\reservedcharson
\comment{ Must have ONLY ONE of these... trust these macros, they work
for CV:

for papers, etc:
}

\reservedcharsoff

\if\papercls y

\else

\fi

\reservedcharson
\comment{Use this around text changed in each revision:}

\comment{Put the next one of these at the top of the paper file after
  each revision so that only the current changes to be reviewed show.

}
\reservedcharsoff

\if\papercls y
\comment{
\received{}
\revised{}
\accepted{}
}
\slugcomment{\tablebox{
In preparation for {\em ApJ}. DRAFT of {\today}.}}
\else
\slugcomment{In preparation for {\em ApJ}.  DRAFT of {\today} \now.}
\fi

\comment{
There may be lots more header in the document, so turn all the
active characters on.  
We (and bibtex) use the active character ~ in names and a few other
places, and we use \sim rather than ~ when wishing for that %
symbol, so turn the active character on before the \begin{document}
even if the others are off with:
\actcharon
}
\reservedcharson
\newcommand\patricio[1]{{\color{cyan} #1}}
\usepackage{textcase}

\comment{The \lowercase call here doesn't make it lowercase, it makes it small caps}
\shorttitle{On the Dayside Atmosphere of WASP-12\lowercase{b}}

\shortauthors{Himes and Harrington}

\begin{document}
\reservedcharson

\comment{But here, it works. Unsure how to fix.}
\title{On the Dayside Atmosphere of WASP-12\lowercase{b}}

\author{
Michael D.\ Himes\altaffilmark{1} and
Joseph Harrington\altaffilmark{1}
}

\affil{\sp{1} Planetary Sciences Group, Department of Physics,
       University of Central Florida, Orlando, FL 32816-2385}
\email{mhimes@knights.ucf.edu}

\begin{abstract}

The atmospheric structure of WASP-12b has been hotly contested for years, with disagreements on the presence of a thermal inversion as well as the carbon-to-oxygen ratio, C/O, due to retrieved abundances of H$_2$O, CO$_2$, and other included species such as HCN and C$_2$H$_2$. Previously, these difficult-to-diagnose discrepancies have been attributed to model differences; assumptions in these models were thought to drive retrievals toward different answers. Here, we show that some of these differences are independent of model assumptions and are instead due to subtle differences in the inputs, such as the eclipse depths and line-list databases. We replicate previously published retrievals and find that the retrieved results are data driven and are mostly unaffected by the addition of species such as HCN and C$_2$H$_2$. We also propose a new physically motivated model that takes into consideration the formation of H$^{-}$ via the thermal dissociation of H$_2$O and H$_2$ at the temperatures reached in the dayside atmosphere of WASP-12b, but the data's current resolution does not support its inclusion in the atmospheric model.  This study raises the concern that other exoplanet retrievals may be similarly sensitive to slight changes in the input data.

\end{abstract}

\keywords{WASP-12b ---
          methods: statistical ---
          planetary systems   ---
          techniques: retrieval}

\section{INTRODUCTION}
\label{sec:introduction}

The thousands of exoplanets discovered to date span a wide range of properties and conditions, from small, rocky bodies to hot, Jupiter-like gas giants \citep{Batalha2014nasKeplerPopulation, WinnFabrycky2015anrevaaExoplanetPopulation}.  Understanding the compositions of these planets provides real-world tests for atmospheric simulations and formation theories.  Characterizing exoplanetary atmospheres requires an observed spectrum.  Presently, most exoplanets can only be characterized via transit (when the planet moves in front of its host star, as seen from Earth) measurements.  These observations capture starlight that has filtered through the exoplanet's atmosphere at the day--night terminator, imprinting information about its composition.  For hot exoplanets, its secondary eclipse (when the planet moves behind the star, as seen from Earth) can be observed, which measures the planet's thermal emission.  This provides better data than transits to constrain the atmospheric properties \citep{DemingSeager2017jgrExoplanetsReview}.

The inference of atmospheric conditions from observed spectra is known as atmospheric retrieval \citep{Madhusudhan2018bookRetrieval}.  For exoplanets, atmospheric retrieval involves the proposal of atmospheric models from some prior distribution (e.g., uniform or Gaussian), computation of the theoretical observed spectra, and determination of how well the proposed models explain the observations.  Unlike solar system observations which only require least-squares minimization, a Bayesian approach is better suited to estimating the uncertainties of exoplanet retrievals due to the high relative noise levels.  The Bayesian sampler explores the parameter space and accepts/rejects new models with some probability based in part on the goodness of fit.  The collection of accepted models forms the posterior distribution, which informs the range and relative likelihood of values for the model parameters.

WASP-12b stands out as one of the hottest exoplanets found to date. 
It orbits an F9V star with a temperature of 6360$\pm_{140}^{130}$ K at 0.02340$\pm_{0.00050}^{0.00056}$ AU every ${\sim}1.09$ days  \citep{HebbEtal2009WASP12b, CollinsEtal2017ajTTVWASP12b}. 
With a mass of 1.47$\pm_{0.069}^{0.076}$ $M_{J}$ and radius of 1.90$\pm_{0.055}^{0.057}$ $R\sb{J}$, its density of 0.0.266$\pm_{0.014}^{0.015}$ g cm$^{-3}$ is less than a quarter of Jupiter's density of 1.326 g/cm$^{3}$.
Due to its extreme equilibrium temperature ({\textgreater}2500 K), it is expected to be in thermochemical equilibrium \citep{Moses2014rsptaChemicalKinetics}. 

WASP-12b has been the target of numerous observations and analyses since its discovery in 2008.  
Its secondary eclipse has been observed across the near- and mid-infrared by a variety of instruments, including the Hubble Space Telescope (HST) Wide Field Camera 3 (WFC3), Spitzer Space Telescope Infrared Array Camera (IRAC), Canada-France-Hawaii Telescope (CFHT) Wideband Imaging Camera (WIC), Michigan-Dartmouth-MIT (MDM) Observatory TIFKAM, and Apache Point Observatory (APO) Near-Infrared Camera 
\citep[][]{LopezMoralesEtal2010apjlWASP12b, 
CampoEtal2011apjWASP12b, 
CrollEtal2011ajWASP12b, 
ZhaoEtal2012apjWASP12bKsband, 
CowanEtal2012apjWASP12b, 
CrossfieldEtal2012apjWASP12b, 
SwainEtal2013icarusWASP12bHST, 
FohringEtal2013mnrasWASP12b}. 
Combinations of these data have been used for retrievals of the dayside $T(p)$ pressure--temperature profile and molecular abundances by 
\citet{MadhusudhanEtal2011natWASP12batm, 
LineEtal2014apjRetrievalCO, 
StevensonEtal2014apjWASP12b}, and 
\citet{OreshenkoEtal2017apjlWASP12b} 
to investigate the atmospheric properties of this highly irradiated hot Jupiter. 
\citet{MadhusudhanEtal2011natWASP12batm} uses less data than the others due to the limited data at the time of publication. 
Further, that retrieval occurred before the discovery of WASP-12's binary M-dwarf companions, which reduces the measured eclipse depths \citep{BergforsEtal2011iauWASP12companion, BechterEtal2014WASP12BC}. 
As a result, our paper does not thoroughly compare those results to other investigations. 

The results of the \citet{LineEtal2014apjRetrievalCO}, \citet{StevensonEtal2014apjWASP12b}, and \citet{OreshenkoEtal2017apjlWASP12b} retrievals are inconsistent in some respects. When considering CO, CO$_2$, CH$_4$, and H$_2$O, \citet{LineEtal2014apjRetrievalCO} do not find evidence for a high C/O due to high abundances of CO$_2$ and H$_2$O, while \citet{StevensonEtal2014apjWASP12b} find a bimodal C/O, both of which have a high abundance of CO$_2$. However, these analyses find an abundance of CO$_2$ that is greater than both CO and H$_2$O, which has been shown to be highly improbable in the atmosphere of a planet like WASP-12b \citep{Madhusudhan2012apjCtoORatio, MosesEtal2013apjCtoORatio, HengLyons2016apjH2AtmospheresCarbonSpecies}. \citet{LineEtal2014apjRetrievalCO} comment that this is implausible and place an upper limit of 10${-5}$ on the CO$_2$ mixing ratio; retrievals under this limit drive the H$_2$O mixing ratio over 100 parts per million, resulting in a more realistic CO$_2$ mixing ratio and maintaining a C/O near solar. \citet{StevensonEtal2014apjWASP12b} also mention the implausibility of their retrieved CO$_2$ abundance and propose the addition of HCN and C$_2$H$_2$ into the retrieval model to solve this problem. These species have been shown to exist when C/O $> 1$ \citep{Madhusudhan2012apjCtoORatio, MosesEtal2013apjCtoORatio} and have spectral features in Spitzer's IRAC channel 2; this allows the Bayesian sampler to fit the eclipse depths in that channel using the added species. Consequently, they retrieve an abundance of CO$_2$ that is less than CO, which is a physically plausible result. This C-rich result is more probable than their O-rich result by a factor of 670. They exclude an isothermal model at $7\sigma$ significance.

\citet{OreshenkoEtal2017apjlWASP12b} performed retrievals using the same data as \citet{StevensonEtal2014apjWASP12b} and expands upon that work by including clouds in their model. They find that the cloud compositions are unconstrained, an expected result considering the degeneracy between cloud composition and gas mixing ratios at low spectral resolutions.  When considering CO, CO$_2$, CH$_4$, and H$_2$O, they replicate the results of \citet{LineEtal2014apjRetrievalCO} and \citet{StevensonEtal2014apjWASP12b} of an unrealistically high CO$_2$ mixing ratio.  In general, they find that their retrieval results are prior dominated.  When assuming Gaussian priors for the C/H and O/H ratios of WASP-12b matching that of its host star \citep{TeskeEtal2014apjStarsCORatio}, the resulting C/O is close to solar.  However, when increasing the uncertainties on the values, C/O $> 1$ becomes the favored solution, consistent with \citet{MadhusudhanEtal2011natWASP12batm}.  Including HCN and C$_2$H$_2$ in the model results in an HCN mixing ratio of $10^{-2}$ -- $10^{-1}$, which they comment is implausible for reasons discussed in \citet{MosesEtal2013apjCtoORatio}; this mixing ratio is over three orders of magnitude greater than that found in \citet{StevensonEtal2014apjWASP12b}.  An interesting result of their retrieval models is the wide range of possible $T(p)$ profiles based on the assumed prior, ranging from no to strong inversions. However, despite this wide range of possibilities, the resulting spectra are qualitatively similar.

A curious difference among the retrieved $T(p)$ profiles of these three investigations is that \citet{LineEtal2014apjRetrievalCO} find the temperature of the atmosphere to be almost entirely above 3000 K with the potential for an inversion, while \citet{StevensonEtal2014apjWASP12b} and \citet{OreshenkoEtal2017apjlWASP12b} find an upper limit of 3000 K with no inversion. 
Note that when computing opacities using HITRAN databases, the available data to calculate partition functions has an upper limit of 3000 K \citep{LaraiaEtal2011icarusTIPS}. The CHIMERA code used by \citet{LineEtal2014apjRetrievalCO} probes temperatures above this limit by assuming that cross sections at temperatures $>3000$ K are equal to those at 3000 K (M. Line, priv. comm.). In general, cross sections will differ between temperatures $>3000$ K and 3000 K, so this assumption is more likely to give misleading results the greater the deviation from 3000 K. However, extrapolation beyond 3000 K could be even more misleading as there is no guarantee that it will match the true cross sections. This assumption in CHIMERA could lead to a better model fit by allowing the code to explore background emission temperatures greater than 3000 K. Nonetheless, this underscores the need for higher-temperature data to more accurately characterize planets in this temperature regime. 

The temperatures found in these retrievals bring attention to another implicit assumption put into these models. Retrieval models consider some set of molecules to fit an observed spectrum. Omitting a molecule that is present in the real object will therefore bias the results: \citet{StevensonEtal2014apjWASP12b} showed that the omission of HCN and C$_2$H$_2$ drives up the inferred CO$_2$ abundance, while including the additional molecules allows for a more realistic fit. At the temperatures retrieved for WASP-12b, both H$_2$ and H$_2$O thermally dissociate, forming H \citep{ArcangeliEtal2018apjWASP-18bNegH, KreidbergEtal2018ajWASP103b, ParmentierEtal2018aapWASP121b}. Some H gain an electron from ionized metals, forming H$^{-}$. To date, WASP-12b retrieval models have omitted H$^{-}$, which provides an important continuum opacity source from bound-free and free-free transitions \citep{John1988aapNegHydrogen}, as more thoroughly discussed in \citet{ParmentierEtal2018aapWASP121b} and \citet{ArcangeliEtal2018apjWASP-18bNegH}.

In this paper, we perform retrievals using the Bayesian Atmospheric Radiative Transfer code  \citep[BART,][]{HarringtonEtal2020apjsBART1, CubillosEtal2020apjsBART2, BlecicEtal2020apjsBART3} matching the setups of \citet{LineEtal2014apjRetrievalCO} and \citet{StevensonEtal2014apjWASP12b} to investigate the discrepancies in their results, and we present a new, physically motivated model that includes additional species not considered in previous investigations. Section \ref{sec:bart} describes the BART code, and Section \ref{sec:setups} discusses the setup for each of the nine models. In Section \ref{sec:resultsdiscussion}, we discuss our results in the context of previous analyses of WASP-12b's dayside atmosphere. Finally, we draw conclusions from our findings in Section \ref{sec:conclusions}.

\section{BART}
\label{sec:bart}

Our retrieval code, BART \citep{HarringtonEtal2020apjsBART1, CubillosEtal2020apjsBART2, BlecicEtal2020apjsBART3}, pairs the Transit radiative-transfer code \citep{Rojo2006PhD} with Multi-Core Markov chain Monte Carlo \citep[MCcubed,][]{CubillosEtal2017apjRednoise}, a Bayesian framework.  The user specifies a parameter space to be explored for some model parameters (e.g., the $T(p)$ profile and molecular abundances).  Other inputs include the observational data and its type (e.g., transit or eclipse depths) as well as the instrument filters associated with each data point.  For each proposed atmospheric model, the theoretical spectrum is calculated at a high resolution, binned according to the filters, and compared to the observational data.  The abundance profiles begin from a user-specified atmosphere (e.g., uniform profiles, or thermochemical equilibrium for a certain $T(p)$ profile), and MCcubed scales the abundances of the molecules being fit.  For credible regions estimated from the posterior, BART computes the steps per effective independent sample (SPEIS) and effective sample size (ESS) to estimate the uncertainty in a given credible region, as detailed in \citet{HarringtonEtal2020apjsBART1}.

\section{MODEL CONFIGURATIONS}
\label{sec:setups}

\atabon\begin{table*}[!ht]
\centering
\caption{Summary of Retrieval Models}
\label{tbl:setups}
\begin{tabular}{cccccccccccccc}
\hline\hline
Characteristic   & 1 & 2 & 3 & 4 & 5 & 6 & 7 & 8 & 9 & 10 & 11 & 12 & 13\\
\hline
CO & \checkmark & \checkmark & \checkmark & \checkmark & \checkmark & \checkmark & \checkmark & \checkmark & \checkmark & \checkmark & \checkmark & \checkmark & \checkmark
\\
CO$_2$ & \checkmark & \checkmark & \checkmark & \checkmark & \checkmark & \checkmark & \checkmark & \checkmark & \checkmark & \checkmark & \checkmark & \checkmark & \checkmark
\\
CH$_4$ & \checkmark & \checkmark & \checkmark & \checkmark & \checkmark & \checkmark & \checkmark & \checkmark & \checkmark & \checkmark & \checkmark & \checkmark & \checkmark 
\\
H$_2$O & \checkmark & \checkmark & \checkmark & \checkmark & \checkmark & \checkmark & \checkmark & \checkmark & \checkmark & \checkmark & \checkmark & \checkmark & \checkmark
\\
HCN & & & & & \checkmark & & \checkmark & & \checkmark & & \checkmark & \checkmark & \checkmark
\\
C$_2$H$_2$ & & & & & \checkmark & & \checkmark & & \checkmark & & \checkmark & \checkmark & \checkmark
\\
NH3 & & & & & & & & & & & & \checkmark & \checkmark
\\
TiO & & & & & & & & & & & & \checkmark & \checkmark
\\
H$^{-}$ & & & & & & & & \checkmark & \checkmark & \checkmark & \checkmark & \checkmark & \checkmark
\\
e$^{-}$ & & & & & & & & \checkmark & \checkmark & \checkmark & \checkmark & \checkmark & \checkmark
\\
\makecell{Cross sections \\ $>3000$ K} & \checkmark & \checkmark & \checkmark & & & \checkmark & & \checkmark & \checkmark & \checkmark & \checkmark &  \checkmark & \checkmark 
\\
HST WFC3 G141 & Sw13$^a$ & Sw13 & Sw13 & St14$^b$ & St14 & St14 & St14 & Sw13 & Sw13 & St14 & St14 & Sw13 & St14
\\
APO ARC z'$^c$ & & & \checkmark & \checkmark & \checkmark & \checkmark & \checkmark & \checkmark & \checkmark & \checkmark & \checkmark & \checkmark & \checkmark
\\
CFHT WIC J$^d$ & & & & \checkmark & \checkmark & \checkmark & \checkmark & & & \checkmark & \checkmark & & \checkmark
\\
CFHT WIC H$^d$ & \checkmark & \checkmark & \checkmark & \checkmark & \checkmark & \checkmark & \checkmark & \checkmark & \checkmark & \checkmark & \checkmark & \checkmark & \checkmark
\\
CFHT WIC Ks$^d$ & & & & \checkmark & \checkmark & \checkmark & \checkmark & & & \checkmark & \checkmark & & \checkmark
\\
\makecell{MDM Hiltner \\ TIFKAM Ks$^e$} & \checkmark & \checkmark & \checkmark & & & & & \checkmark & \checkmark & & & \checkmark & 
\\
\makecell{Subaru MOIRCS \\ NB2315$^f$} & \checkmark & \checkmark & \checkmark & & & & & \checkmark & \checkmark & & & \checkmark & \\
Spitzer IRAC ch1 & Co12$^g$ & Co12 & Co12 & St14 & St14 & St14 & St14 & Co12 & Co12 & St14 & St14 & Co12 & St14
\\
Spitzer IRAC ch2 & Co12 (n)$^i$ & Co12 (e)$^j$ & Co12 (n) & St14 & St14 & St14 & St14 & Co12 (n) & Co12 (n) & St14 & St14 & Co12 (n) & St14
\\
Spitzer IRAC ch3 & Ca11$^h$ & Ca11 & Ca11 & St14 & St14 & St14 & St14 & Ca11 & Ca11 & St14 & St14 & Ca11 & St14
\\
Spitzer IRAC ch4 & Ca11 & Ca11 & Ca11 & St14 & St14 & St14 & St14 & Ca11 & Ca11 & St14 & St14 & Ca11 & St14
\\
\hline
\end{tabular}\\
a: \citet{SwainEtal2013icarusWASP12bHST}\\
b: \citet{StevensonEtal2014apjWASP12b}\\
c: \citet{LopezMoralesEtal2010apjlWASP12b}, corrected by \citet{CrossfieldEtal2012apjWASP12b}\\
d: \citet{CrollEtal2011ajWASP12b}, corrected by \citet{CrossfieldEtal2012apjWASP12b}\\
e: \citet{ZhaoEtal2012apjWASP12bKsband}, corrected by \citet{CrossfieldEtal2012apjWASP12b}\\
f: \citet{CrossfieldEtal2012apjWASP12b}\\
g: \citet{CowanEtal2012apjWASP12b}, corrected by \citet{CrossfieldEtal2012apjWASP12b}\\
h: \citet{CampoEtal2011apjWASP12b}, corrected by \citet{CrossfieldEtal2012apjWASP12b}\\
i: `null' hypothesis\\
j: `ellipsoidal' hypothesis
\end{table*}\ataboff

To investigate the previously published retrievals of WASP-12b, we replicate their setups to the best of BART's ability, and we expand upon those setups to delve deeper into the nature of the discrepancies in results. Our nine retrieval models are that of 

\begin{enumerate}
    \item \citet{LineEtal2014apjRetrievalCO} \textit{null} case,
    \item \citet{LineEtal2014apjRetrievalCO} \textit{ellipsoidal} case,
    \item \citet{LineEtal2014apjRetrievalCO} \textit{null} case plus the data from \citet{LopezMoralesEtal2010apjlWASP12b} corrected by \citet{CrossfieldEtal2012apjWASP12b},
    \item \citet{StevensonEtal2014apjWASP12b} case without HCN and C$_2$H$_2$,
    \item \citet{StevensonEtal2014apjWASP12b} case with HCN and C$_2$H$_2$,
    \item \citet{StevensonEtal2014apjWASP12b} without HCN and C$_2$H$_2$ with the assumptions of CHIMERA about cross sections,
    \item \citet{StevensonEtal2014apjWASP12b} with HCN and C$_2$H$_2$ mixing ratios fixed to their reported C-rich best-fit values,
    \item Model 3, with H$^{-}$,
    \item Model 8, with HCN and C$_2$H$_2$,
    \item Model 6, with H$^{-}$, 
    \item Model 10, with HCN and C$_2$H$_2$, 
    \item Model 3, with NH$_3$, HCN, C$_2$H$_2$, TiO, and H$^{-}$, and
    \item Model 6, with the molecules of Model 12.
\end{enumerate}

At the time of writing, BART does not have a realistic cloud model, so we do not try to replicate the \cite{OreshenkoEtal2017apjlWASP12b} result that finds cloud composition to be unconstrained. Table \ref{tbl:setups} summarizes the setup of each model regarding data sources and molecules besides H, H$_2$, and He.

For this investigation, BART's only restrictions on the atmospheric models are 1) the sum of molecular abundances must equal 1 for each layer, and 2) the ratio of H$_2$ to He is held constant by adjusting their abundances to satisfy condition 1.  
For models that do not use CHIMERA's assumption about cross sections at $T > 3000$ K, BART also enforces that the temperature of the atmosphere must remain within the line-list limits.

The atmospheric models consist of 100 log-spaced layers spanning 10$^{-8}$ -- 100 bar. For radiative-transfer calculations, only layers above where the optical depth reaches ${\geq}10$ are considered.  We assume uniform abundances for the species present, consistent with the previous publications.  Each model has a free parameter for the abundance of each opacity-contributing species, as well as five free parameters for the $T(p)$ profile \citep[the Planck mean infrared opacity, the ratios of the Planck mean visible and infrared opacities for two streams, the partition between the two streams, and a general parameter for albedo/emissivity/energy recirculation; ][]{LineEtal2013apjRetrieval1}.  The free parameters for the partition between streams and albedo/emissivity/recirculation have uniform priors; all other free parameters have log-uniform priors.  We do not vary the H$^{-}$ or e$^{-}$ abundances because previous publications indicate the atmosphere is nearly isothermal at ${\sim}3000$ K in the regions with sensitivity, and the abundances change by only a factor of ${\sim}2$ for a change as large as 200 K. H$^{-}$ and e$^{-}$ are fixed to an abundance of $10^{-9}$ and $10^{-6}$, respectively, which is roughly consistent with thermochemical equilibrium at 3000 K \citep[NASA Chemical Equilibrium with Applications code][]{GordonMcBride1994reportCEA} at pressures probed by the observations.  A wide range of values are allowed for each free parameter without consideration of physical plausibility. For this investigation, we use the DEMCzs sampling algorithm of \citet{Braak2008SnookerDEMC} because we found that the DEMC algorithm of \citet{Braak2006DifferentialEvolution} occasionally leads to rogue chains that do not converge.  Since DEMCzs only considers the goodness of fit of each proposed model, it can explore both realistic and unrealistic solutions. The initial samples of parameters for the DEMCzs algorithm are drawn randomly from a uniform distribution; most of the parameters are the logarithm of the true parameter, so those true parameters are randomly sampled from a log-uniform space.

We include HITEMP opacities for CO, CO$_2$, and H$_2$O \citep{RothmanEtal2010jqsrtHITEMP}, and HITRAN opacities for NH$_3$, C$_2$H$_2$, and HCN \citep{Rothman2013jqsrtHITRAN2012}.  Models 1 -- 11 use the \citet{Rothman2013jqsrtHITRAN2012} CH$_4$ line list for consistency with \citet{LineEtal2014apjRetrievalCO}.   while models 12 and 13 use the new CH$_4$ HITEMP line list \citep{HargreavesEtal2020arxivMethaneHITEMP}.  TiO opacities are sourced from \citet{Schwenke1998fdTiO}.  We include H$_2$-H$_2$ and H$_2$-He collision-induced absorptions \citep{RichardEtal2012jqsrtHITRANcia} as well as H$^{-}$ bound-free and free-free absorption \citep{John1988aapNegHydrogen}, where appropriate. 

Note that Equation 3 of \citet{John1988aapNegHydrogen} does not lead to the correct bound-free opacity values necessary to reproduce Table 1 of \citet{John1988aapNegHydrogen}; a factor of 10 for the bound-free cross-sections is required to obtain agreement. We refer the reader to our compendium for a detailed proof. It is unclear whether the given equations or table provide the correct opacities. Considering the numerous fitted constants used in the equations, we have chosen to assume that the table is correct, as it appears to lead to agreement with the H$^{-}$ opacities plotted in Figure 1 of \citet{ArcangeliEtal2018apjWASP-18bNegH} and Figure 4 of \citet{ParmentierEtal2018aapWASP121b}.  Additionally, by assuming the greater of the two possibilities, we can assess an upper limit on the impact of H$^{-}$ when retrieving WASP-12b's atmospheric properties.

\section{RESULTS & DISCUSSION}
\label{sec:resultsdiscussion}

\begin{figure*}[ht]
\includegraphics{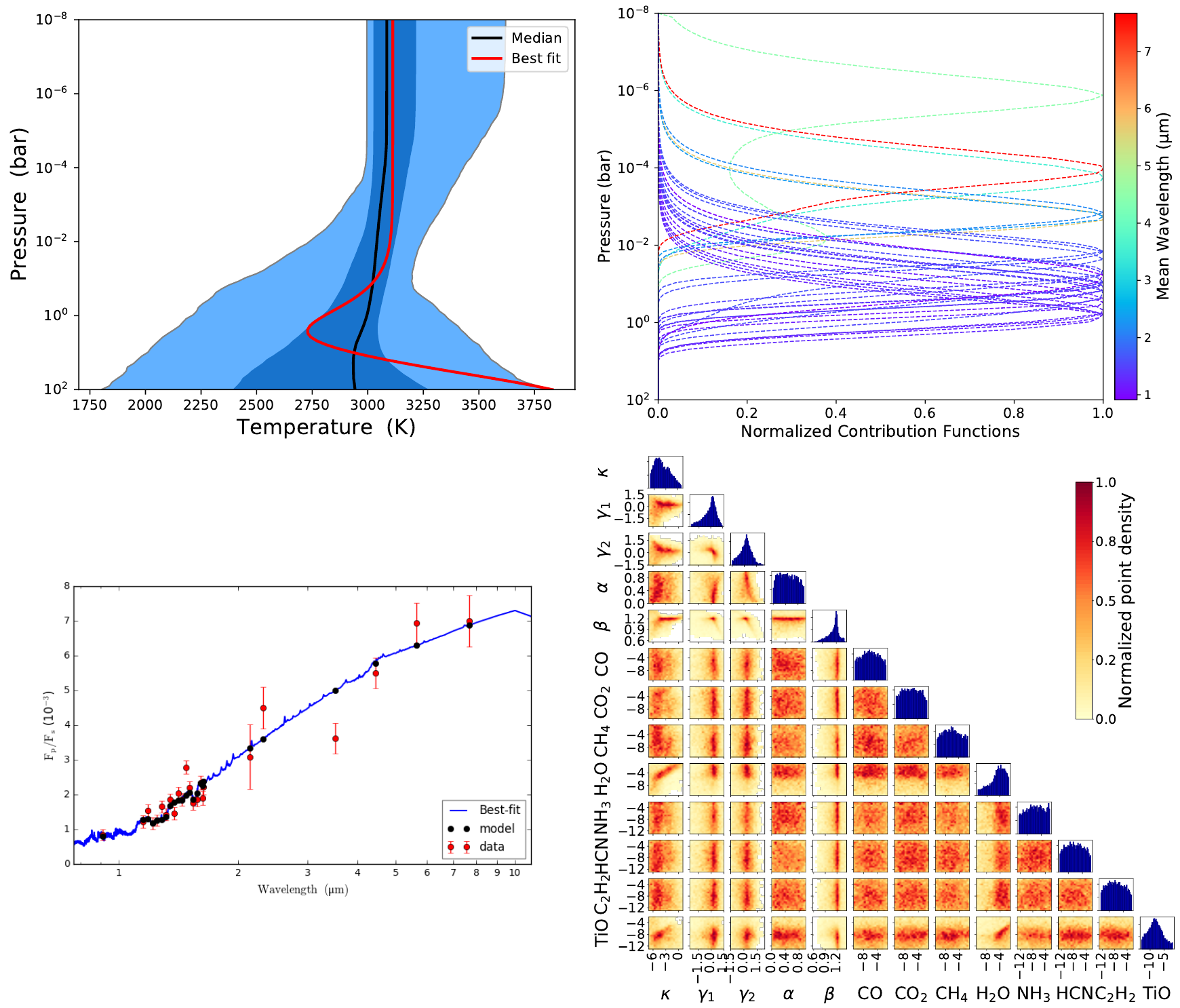}
\caption{BART results for Model 12, which uses the data set of \citet{LineEtal2014apjRetrievalCO} with the additional data point from \citet{LopezMoralesEtal2010apjlWASP12b}.  \textbf{Top left:} $T(p)$ profiles explored by the MCMC. Red line denotes the best-fit $T(p)$ profile, the black line denotes the median $T(p)$ profile, and the dark and light blue regions indicate the $1{\sigma}$ and $2{\sigma}$ regions, respectively. \textbf{Top right:} normalized contribution functions. \textbf{Bottom left:} best-fit spectrum.  \textbf{Bottom right:} Marginalized posteriors.  BART favors a mostly-isothermal $T(p)$ profile around 3000 K.  Of the molecules considered, only H$_2$O is constrained (log abundance \textgreater -6).  While the log abundance of TiO features a bump near -8 -- -7, the non-negligible tail at lesser abundances indicates the possibility that TiO is not present in the atmosphere.
\label{fig:model12}}
\end{figure*}

\begin{figure*}[ht]
\includegraphics{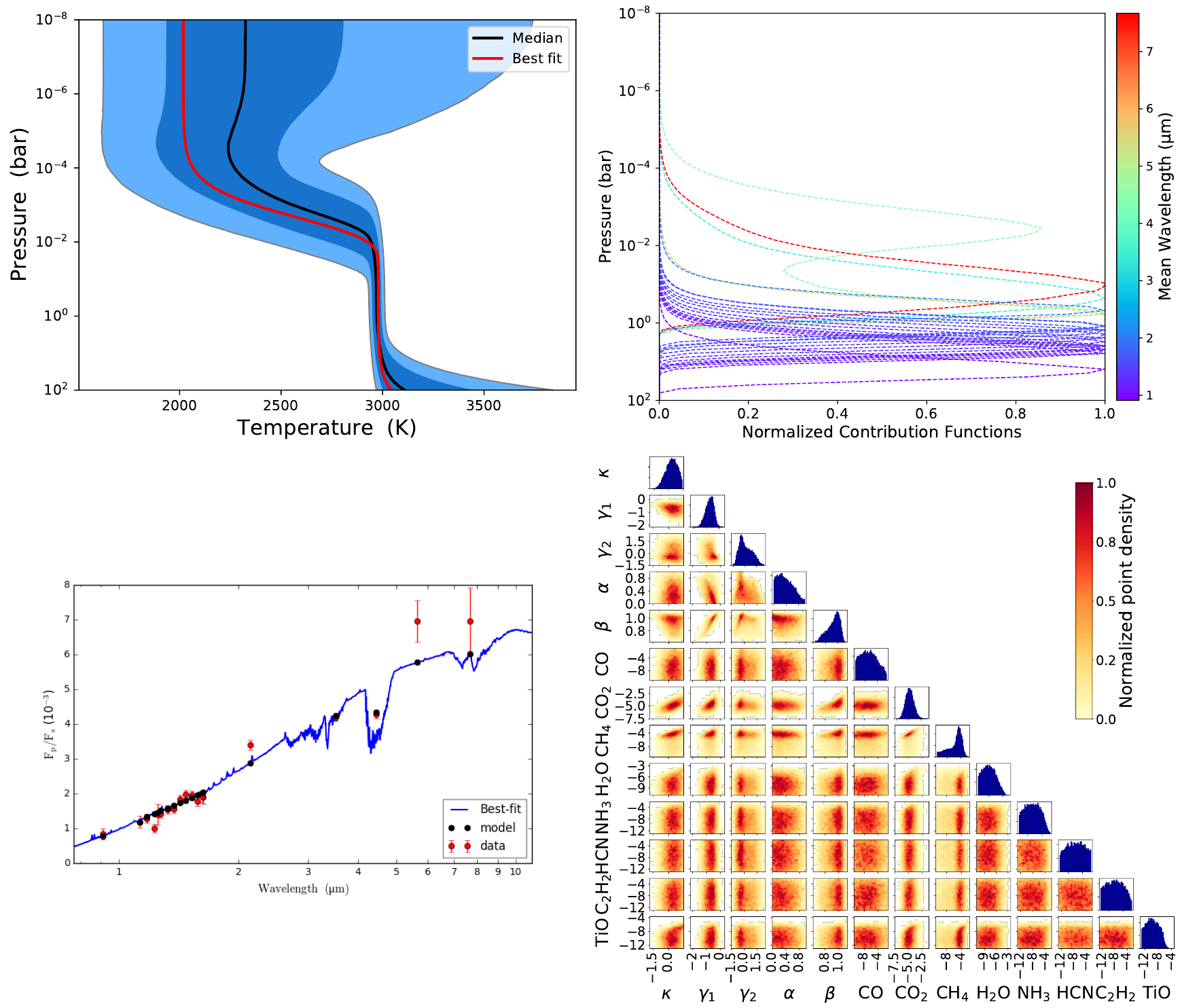}
\caption{Same as Figure \ref{fig:model12}, except for Model 13, which uses the data set of \citet{StevensonEtal2014apjWASP12b}.  Like Model 12, BART favors a nearly isothermal $T(p)$ profile in the regions probed by the observations.  Contrary to Model 12, H$_2$O features an upper limit around -4 for its log abundance, CO$_2$ is constrained to a log abundance of -7 -- -3, CH$_2$ shows evidence of a log abundance around -5 -- -4 (though features a non-negligible tail at lesser abundances, indicating it may not be present in the atmosphere), and CO favors log abundances around -10 -- -6.  Like Model 12, the posterior for TiO indicates an upper limit on the log abundance of around -5 -- -4, and there is similarly no evidence for HCN or C$_2$H$_2$.
\label{fig:model13}}
\end{figure*}

The accepted $T(p)$ profiles with $1\sigma$ and $2\sigma$ regions, normalized contribution functions, best-fit spectrum, and 1D marginalized posteriors are shown for Models 12 and 13 in Figure \ref{fig:model12} and \ref{fig:model13}, respectively.  Appendix \ref{sec:appwaspfigs} provides additional figures: 2D pairwise posteriors and trace plots for Models 12 and 13, as well as the corresponding set of six plots for Models 1 -- 11.  Table \ref{tbl:retrievedabundances} contains the best-fit values and 68.27\% interval for the retrieved log abundances of each molecule for all 13 models, the best-fit values and 68.27\% interval reported by \citet{LineEtal2014apjRetrievalCO} for the `null' and `ellipsoidal' cases, the best-fit values reported by  \citet{StevensonEtal2014apjWASP12b} for the C-rich and O-rich cases, and the lower/upper limits on abundances shown in the top of Figure 3 of \citet{OreshenkoEtal2017apjlWASP12b}. We also report the best-fit and  68.27\% credible region for C/O for each case.  We have excluded extreme outliers (C/O $> 1000$) from the density estimation, as they represent a small percentage of the total models and cause problems for the density estimation algorithm, and we do not consider HCN or C$_2$H$_2$ when calculating C/O due to the lack of evidence to support their inclusion in the model.  Values of ``\ldots" indicates that the model does not contain that molecule.  Table \ref{tbl:speis-ess} lists the SPEIS, ESS, and associated uncertainty in the 68.27\% credible region for each model considered.  We also provide a compendium with the data and commands necessary to reproduce this work; the link is at the end of the text.

\atabon\begin{table*}[ht]
\centering
\caption{Retrieved Molecular Log Abundances\label{tbl:retrievedabundances}}
\begin{tabular}{cccccccccc}
\hline\hline
Model & CO & CO\sb{2} & CH\sb{4} & H\sb{2}O & HCN & C\sb{2}H\sb{2} & NH\sb{3} & TiO & C/O \\
\hline
1     & -4.4 & -2.6 & -11.4 & -5.4 & \ldots & \ldots & \ldots & \ldots & 0.50  \\
      & [-10.0, -2.3] & [-11.0, -2.2] & [-10.5, -2.9] & [-4.7, -1.2] & \ldots & \ldots & \ldots & \ldots & [0.00, 4.81] \\
2     & -11.3 & -5.2 & -2.8 & -11.1 & \ldots & \ldots & \ldots & \ldots & 115  \\
      & [-10.9, -4.7] & [-5.6, -3.2] & [-9.7, -2.0] & [-11.2, -6.6] & \ldots & \ldots & \ldots & \ldots & [0.00, 9.53] \\
3     & -2.0 & -1.0 & -10.1 & -6.6 & \ldots & \ldots & \ldots & \ldots & 0.52  \\
      & [-9.6, -2.1] & [-10.8, -1.4] & [-11.2, -4.2] & [-5.5, -1.2] & \ldots & \ldots & \ldots & \ldots & [0.00, 4.84 \\
4     & -10.4 & -5.8 & -3.8 & -8.6 & \ldots & \ldots & \ldots & \ldots & 52.8  \\
      & [-13.2, -5.7] & [-6.2, -5.0] & [-6.6, -2.8] & $< -7.7$ & \ldots & \ldots & \ldots & \ldots & [0.06, 9.86] \\
5     & -6.8 & -5.5 & -3.5 & -8.3 & -7.0 & -7.1 & \ldots & \ldots &  52.2 \\
      & [-10.6, -4.8] & [-6.1, -4.9] & [-6.1, -2.6] & [-10.6, -6.7] & [-10.6, -1.4] & [-10.0, -4.1] & \ldots & \ldots & [0.00, 12.4] \\
6     & -5.9 & -5.2 & -3.2 & -8.4 & \ldots & \ldots & \ldots & \ldots & 45.2  \\
      & [-10.9, -4.6] & [-5.5, -4.0] & [-5.6, -1.6] & [-11.1, -6.6] & \ldots & \ldots & \ldots & \ldots & [0.10, 12.6] \\
7     & -7.8 & -5.5 & -3.5 & -7.2 & -6 & -5 & \ldots & \ldots &  50.5 \\
      & [-11.3, -5.5] & [-6.1, -4.9] & [-7.3, -2.7] & [-10.8, -6.9] & fixed & fixed & \ldots & \ldots & [0.04, 23.7] \\
8     & -2.3 & -1.0 & -5.7 & -7.3 & \ldots & \ldots & \ldots & \ldots & 0.51 \\
      & [-8.2, -1.3] & [-10.0, -1.6] & [-10.5, -3.4] & [-6.8, -1.0] & \ldots & \ldots & \ldots & \ldots & [0.00, 5.72] \\
9     & -2.2 & -1.7 & -9.4 & -8.4 & -8.5 & -2.8 & \ldots & \ldots & 0.57 \\
      & [-10.3, -1.5] & [-10.1, -1.3] & [-10.2, -2.5] & [-6.6, -1.1] & [-12.1, -2.4] & [-11.9, -3.6] & \ldots & \ldots & [0.00, 5.64] \\
10    & -6.7 & -5.5 & -3.7 & -10.0 & \ldots & \ldots & \ldots & \ldots & 37.3 \\
      & [-10.5, -4.9] & [-6.1, -4.5] & [-5.7, -1.9] & [-10.7, -6.5] & \ldots & \ldots & \ldots & \ldots & [0.05, 21.0] \\
11    & -8.4 & -5.9 & -3.9 & -10.9 & -10.5 & -6.3 & \ldots & \ldots & 45.9 \\
      & [-10.8, -5.2] & [-6.0, -4.4] & [-6.3, -1.9] & [-10.4, -6.6] & [-11.6, -3.5] & [-12.6, -5.3] & \ldots & \ldots & [0.05, 20.0] \\
12    & -9.5 & -2.1 & -1.6 & -8.0 & -9.4 & -12.2 & -11.5 & -8.1 & 1.94  \\
      & [-8.3, -2.1] & [-9.6, -2.3] & [-10.5, -2.5] & [-4.5, -1.2] & [-11.8, -3.7] & [-11.0, -4.9] & [-10.6, -4.9] & [-10.0, -5.1] & [0.00, 5.51] \\
13    & -9.1 & -5.2 & -4.4 & -8.1 & -5.4 & -6.0 & -8.2 & -9.8 & 3.39  \\
      & [-10.2, -4.8] & [-5.8, -4.2] & [-6.8, -3.2] & [-10.0, -6.0] & [-11.2, -3.2] & [-12.0, -4.8] & [-12.4, -6.4] & [-12.3, -8.4] & [0.31, 1.43] \\
\hline
L14$^a$ null & -2.7 & -1.2 & -8.6 & -8.1 & \ldots & \ldots & \ldots & \ldots & 0.51 \\
  & [-9.7, -2.0] & [-8.1, -1.3] & [-10.8, -3.1] & [-8.8, -2.3] & \ldots & \ldots & \ldots & \ldots & [0.30, 1.00] \\
L14 ellipsoidal & -2.7 & -1.0 & -9.7 & -3.3 & \ldots & \ldots & \ldots & \ldots & 0.50 \\
 & [-10.3, -2.8] & [-1.3, -0.8] & [-10.6, -5.1] & [-10.0, -3.1] & \ldots & \ldots & \ldots & \ldots & [0.11, 0.22] \\
St14 C-rich & -3.5 & -6.0 & -4.1 & -6.6 & -6.0 & -5.0 & \ldots & \ldots & 1.30 \\
St14 O-rich & -3.3 & -4.2 & -7.0 & -3.3 & -7.0 & -9.8 & \ldots & \ldots & 0.50 \\
O17$^b$ & [-12, -2.5] & [-7, -4] & [-12, -6] & [-12, -6] & [-3, -1] & [-12, -4] & \ldots & \ldots & [0.3, 4] \\
\hline
\end{tabular}\\
\textbf{Notes:} Models 1 -- 13 and the \citet{LineEtal2014apjRetrievalCO} results are given as the best-fit values, followed by the 68.27\% credible regions. \citet{StevensonEtal2014apjWASP12b} results are given as the reported best-fit values. \citet{OreshenkoEtal2017apjlWASP12b} results are given as the estimated minimum and maximum of the posteriors from their Figure 3.\\
a: \citet{LineEtal2014apjRetrievalCO}\\
b: \citet{OreshenkoEtal2017apjlWASP12b}
\end{table*}\ataboff

\atabon\begin{table}[ht]
\centering
\caption{}
\label{tbl:speis-ess}
\begin{tabular}{cccc}
\hline\hline
Model & SPEIS$^a$ & ESS$^b$ & 68.27\% Region Uncertainty (\%)$^c$ \\
\hline
1     & 1834 & 517 & 2.04 \\
2     & 2276 & 417 & 2.27 \\
3     & 1025 & 926 & 1.53 \\
4     &  953 & 996 & 1.47 \\
5     & 4871 & 195 & 3.31 \\
6     & 1948 & 487 & 2.10 \\
7     & 3463 & 274 & 2.80 \\
8     & 5483 & 455 & 2.17 \\
9     & 6135 & 407 & 2.30 \\
10    & 4673 & 534 & 2.01 \\
11    & 7562 & 330 & 2.55 \\
12    & 3356 & 327 & 2.56 \\
13    & 3359 & 327 & 2.56 \\
\hline
\end{tabular}\\
a: Steps per effective independent sample; the number of iterations needed for a non-correlated sample.\\
b: Effective sample size; the number of independent samples.  \\
c: For finite ESS, the approximation to the true posterior induces an uncertainty in any determined credible region.  See \citet{HarringtonEtal2020apjsBART1} for details. \\
\end{table}\ataboff

For Model 1 (Figure \ref{fig:elec-supp-1}), BART's results generally agree with that of the `null' case of \citet{LineEtal2014apjRetrievalCO}. Our best-fit abundances have CO$_2$ $>$ CO, as \citet{LineEtal2014apjRetrievalCO} find, which, as mentioned previously here and in their paper, is implausible.  However, there is large uncertainty in this result, as indicated by the nearly flat posterior of both CO and CO$_2$. Like \citet{LineEtal2014apjRetrievalCO}, we find that CO and CH$_4$ are unconstrained, according to the flat posteriors, but we also find that CO$_2$ is unconstrained. Except for CH$_4$, the best-fit values fall within the 68.27\% region reported by \citet{LineEtal2014apjRetrievalCO}. The $T(p)$ profile $1\sigma$ regions overlap, with both best-fit profiles favoring an inversion.

For Model 2 (Figure \ref{fig:elec-supp-2}), BART's results differ in some respects from the `ellipsoidal' case of \citet{LineEtal2014apjRetrievalCO}. Most notably, the $T(p)$ profile $1\sigma$ region shows a non-inverted atmosphere where the upper atmosphere is $< 3000$ K, whereas \citet{LineEtal2014apjRetrievalCO} find an inverted atmosphere with the upper atmosphere $> 3000$ K. However, the normalized contribution functions indicate minimal sensitivity below a pressure of $10^{-6}$ bar; in the best-constrained region (0.1 -- 10 bar), the retrieved $T(p)$ profiles agree. For abundances, the best-fit values disagree, but, except for CO$_2$, the 68.27\% intervals overlap. However, the case of \citet{LineEtal2014apjRetrievalCO} where CO$_2$ has an upper limit agrees more closely with our retrieved interval. It is unclear why BART does not find the high-CO$_2$ mode found by \citet{LineEtal2014apjRetrievalCO}.

The results for Model 3 (Figure \ref{fig:elec-supp-3}) in general agree with Model 1. CO, CO$_2$, and CH$_4$ are similarly unconstrained, while the 68.27\% region for H$_2$O overlaps (Table \ref{tbl:retrievedabundances}). A notable difference is that Model 3 finds a lower minimum for that region. This is likely from the additional data point of \citet{LopezMoralesEtal2010apjlWASP12b} providing an additional constraint on the background emission.

BART's results for Model 4 (Figure \ref{fig:elec-supp-4}) in many ways match that of Model 2. The $T(p)$ profile $1\sigma$ regions closely match, and the marginalized posteriors exhibit many similarities. Similar to Model 2, BART does not find the unrealistically high CO$_2$ abundance reported by \citet{StevensonEtal2014apjWASP12b} when retrieving with this setup (Table \ref{tbl:retrievedabundances}). Since the parameter space for CO$_2$ extended to a log mixing ratio of -1, these models must have been discarded by BART's sampler. It is uncertain whether this difference can be attributed to the sampling algorithm, the opacity sources, or some other difference in the retrieval algorithm. Nevertheless, the best-fit $T(p)$ profiles of \citet{StevensonEtal2014apjWASP12b} fall within BART's reported $1\sigma$ region where there is sensitivity. Further, their C-rich best-fit values are within BART's 68.27\% regions for all but CO. Our 68.27\% interval for H$_2$O rules out their O-rich result, which features an H$_2$O abundance of $10^{-3.3}$.

The results for Model 5 (Figure \ref{fig:elec-supp-5}) mostly agree with the findings of \citet{StevensonEtal2014apjWASP12b}. The C-rich best-fit values for CO$_2$, CH$_4$, H$_2$O, and C$_2$H$_2$ reported by them fall within BART's 68.27\% regions (Table \ref{tbl:retrievedabundances}). Our 68.27\% interval for H$_2$O similarly rules out their O-rich result. BART finds HCN to be unconstrained. The retrieved $T(p)$ profiles closely match the results of Model 4 and are therefore consistent with the retrieved $T(p)$ profiles of \citet{StevensonEtal2014apjWASP12b}. The best-fit values and 68.27\% region found for Model 5 closely matches the results of Model 4, which does not include HCN or C$_2$H$_2$. Thus, the inclusion of these additional molecules only minorly affects the retrieved result.

Model 6 (Figure \ref{fig:elec-supp-6}) examines the effect of allowing $T(p)$ profiles with $T > 3000$ K on Model 4. BART's results closely match those of Models 4 and 5. The $T(p)$ profile $1\sigma$ regions generally agree, with a general upper limit of ${\sim}3000$ K. The marginalized posteriors are similar, and the 68.27\% regions for the molecular abundances closely overlap. This importantly demonstrates that the eclipse data, not the model assumptions, are driving the result.

For Model 7 (Figure \ref{fig:elec-supp-7}), BART's results generally agree with those of Model 5. Even with the HCN and C$_2$H$_2$ abundances fixed to the C-rich best-fit values reported by \citet{StevensonEtal2014apjWASP12b}, BART does not favor their reported CO abundance.  We suspect that this difference can be attributed to differences in line lists, though we were unable to ascertain the line lists used by \citet{StevensonEtal2014apjWASP12b} via private communication to test this hypothesis.

Models 8, 9, and 12 (Figures \ref{fig:elec-supp-8}, \ref{fig:elec-supp-9}, \ref{fig:elec-supp-12}), which are equivalent to Model 3 but with additional opacity sources, bear qualitatively similar results to Model 3.  While the 68\% regions may differ slightly, these are largely due to numerical differences in the density estimation of the marginalized posteriors.  Visually, the posteriors are quite similar; CO, CO$_2$, and CH$_4$ have flat marginalized posteriors, while H$_2$O has a preference for log abundances greater than -6.  In Model 12, the flat posteriors of HCN, C$_2$H$_2$, and NH$_3$ demonstrate that the data are insensitive to their inclusion.  While the log abundance TiO is loosely constrained to be $-7.5\pm{2.5}$, consistent with thermochemical equilibrium for the retrieved $T(p)$ profile in the regions probed by the observations, the posterior allows for TiO to be absent from the atmosphere within $2\sigma$.

Similarly, Models 10, 11, and 13 (Figures \ref{fig:elec-supp-10}, \ref{fig:elec-supp-11}, \ref{fig:elec-supp-13}), which are equivalent to Model 6 except with additional opacity sources, yield results similar to Model 6.  The resulting thermal profiles agree closely.  As with Models 8, 9, and 12, the 68\% regions differ slightly due to the density estimation algorithm, though the marginalized posteriors are qualitatively similar to Model 6.  That is, CO tends to favor smaller values, CO$_2$ is tightly constrained within an order of magnitude from $10^{-5}$, CH$_4$ features a constraint about $10^{-3}$ but could also be absent from the atmosphere (the posterior has a non-negligible tail), and H$_2$O has an upper limit of around $10^{-4}$.  For Model 13, BART finds upper limits of $10^{-2}$ for NH$_3$ and $10^{-6}$ for TiO.  

\begin{figure}[ht]
\includegraphics[width=0.49\textwidth, clip=True]{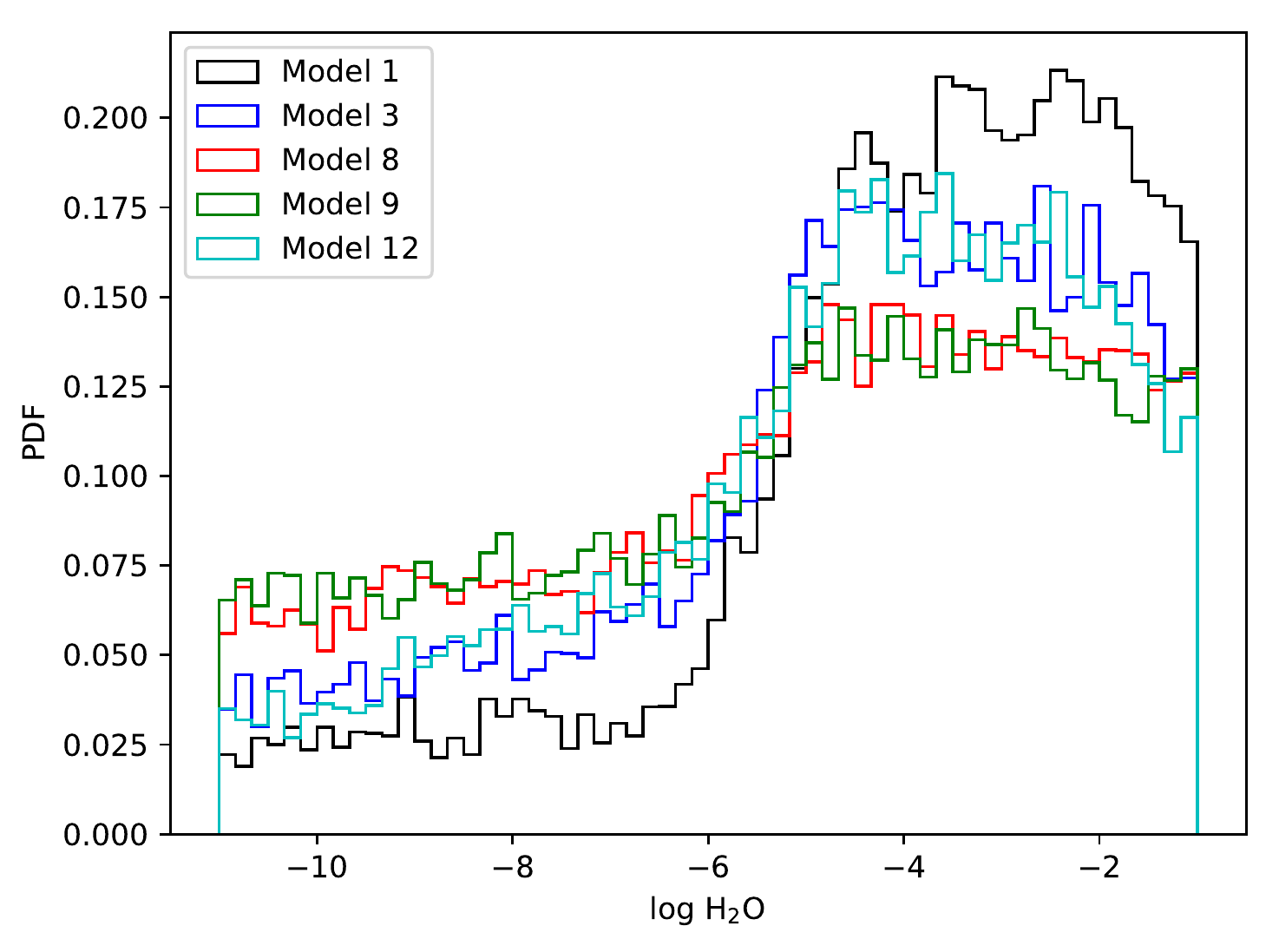}
\caption{Comparison of the retrieved H$_2$O marginalized posteriors for the models using the \citet{LineEtal2014apjRetrievalCO} `null' data set.  The lack of significant variation between the models without H$^{-}$ (1, 3) and those with H$^{-}$ (8, 9, 12) indicates that the inclusion of H$^{-}$ does not significantly affect the retrieved H$_2$O abundance.
\label{fig:compH2O}}
\end{figure}

As Models 8 and 10, 9 and 11, and 12 and 13 bear identical setups/assumptions aside from the eclipse depths, the differences between them are thus solely attributable to the data.  Both data sets favor a ${\sim}3000$ K atmosphere in the regions probed by the observations, as evidenced by the normalized contribution functions.
However, their retrieved abundances, and by extension the inferred C/O, are incompatible.  The data set of Models 8, 9, and 12 yields evidence of a high water abundance and possibly TiO in thermochemical equilibrium, with no evidence of other molecules, while the data set of Models 10, 11, and 13 constrain CO$_2$ and CH$_4$, with upper limits for CO and H$_2$O.  None of the data sets considered offer meaningful constraints on HCN, C$_2$H$_2$, or NH$_3$.  Despite the theoretical expectation that H$^{-}$ plays an important role in the atmosphere of hot Jupiters like WASP-12b \citep{ParmentierEtal2018aapWASP121b}, its inclusion does not make a significant difference in the retrieval results (Figure \ref{fig:compH2O}).  

Our results are broadly consistent with those of \citet{OreshenkoEtal2017apjlWASP12b}, except for HCN (Table \ref{tbl:retrievedabundances}).  In general, the models using the \citet{LineEtal2014apjRetrievalCO} data favor an atmosphere that is isothermal or has an inversion, whereas the models using the \citet{StevensonEtal2014apjWASP12b} data favor an atmosphere that has no inversion. \citet{LothringerEtal2018apjExtremelyIrradiatedHotJupiters} show that thermal inversions are likely for this planet class even without VO or TiO. While the results using the \citet{LineEtal2014apjRetrievalCO} `null' data set are consistent with this finding, it is important to consider that these results also favor a high H$_2$O abundance, which is inconsistent with the expected thermal dissociation of H$_2$O. While the \citet{LineEtal2014apjRetrievalCO} `ellipsoidal' case favors a low H$_2$O abundance, it does not show strong evidence of a thermal inversion and more closely matches the results of \citet{StevensonEtal2014apjWASP12b}.

Over the reported 68.27\% credible regions, the C/O for these models can take a wide range of values.   However, they do not indicate that C/O$\gg$1. Rather, it is due to a combination of reasons.  For one, the DEMCzs sampler is free to explore the parameter space without regard for C/O.  For example, in the case of Model 4, the best-fit C/O value is 52.8, which is noticeably different than the 68.27\% region; this is a product of a high best-fit abundance of CH$_4$ and low best-fit abundance of H$_2$O, which provides a better statistical fit than models with more reasonable C/O values.  Additionally, thermal dissociation of H$_2$O would lead to the formation of O \citep{ParmentierEtal2018aapWASP121b}, causing an apparent increase in the measurable C/O but not the true C/O.  There are also other oxygen-bearing species not considered here, particularly condensates at the limb \citep{WakefordEtal2017mnrasCondensatesWASP12b}, that would contribute to C/O, if included in the model.  However, \citet{OreshenkoEtal2017apjlWASP12b} demonstrated that the cloud composition is degenerate with gas mixing ratios; higher-resolution data is necessary to explore a model with various species of condensates.

Typically, when considering multiple models, the Bayes factor of each model is compared to choose the `best' model.  In this investigation, however, this would be erroneous: the 13 models presented do not use the same data sets, and they are not competing for the `best' model.  Rather, the models demonstrate that the retrieval results are data driven and independent of the model selected.  Consequently, we do not compute the Bayes factor as it would be a misleading metric.  We emphasize that the results show that the previous retrieval analyses of \citet{LineEtal2014apjRetrievalCO} and \citet{StevensonEtal2014apjWASP12b} are consistent, when considering the data set used in each investigation.  Follow-up observations are required to determine which data set, if either, represents the true nature of WASP-12b.

\section{CONCLUSIONS}
\label{sec:conclusions}

In general, we are able to reproduce the published results of \citet{LineEtal2014apjRetrievalCO}, \citet{StevensonEtal2014apjWASP12b}, and \cite{OreshenkoEtal2017apjlWASP12b} using BART. We confirm the finding of \citet{StevensonEtal2014apjWASP12b} that excludes an isothermal profile when mimicking their setup.  By following the model assumption of \citet{LineEtal2014apjRetrievalCO} allowing temperatures above 3000 K with the \citet{StevensonEtal2014apjWASP12b} data, the range of possible $T(p)$ profiles expands to include inverted profiles but still favors a non-inverted atmosphere.  Note that an inverted profile is expected for ultra-hot Jupiters like WASP-12b \citep{LothringerEtal2018apjExtremelyIrradiatedHotJupiters}. 

We find that current data does not support the inclusion of HCN, C$_2$H$_2$,  H$^{-}$, e$^{-}$, NH$_3$, and TiO, as they do not significantly affect the posterior.  As new telescopes in the near future provide higher quality data, these molecules should be reconsidered, as in this investigation, to determine if they inform the results.

Some aspects were unable to be reproduced, namely, the unrealistically high CO$_2$ abundances reported in the \citet{LineEtal2014apjRetrievalCO} `ellipsoidal' case and the \citet{StevensonEtal2014apjWASP12b} case without HCN or C$_2$H$_2$, and the unrealistically high HCN abundance reported by \citet{OreshenkoEtal2017apjlWASP12b}. While we suspect this discrepancy is due to differences in the retrieval model, further investigation is necessary to definitively determine the origin. 

We have demonstrated that differences in eclipse depth data sets primarily drive the differences between the results of \citet{LineEtal2014apjRetrievalCO} and \citet{StevensonEtal2014apjWASP12b}, with more subtle differences likely attributable to the retrieval model and input data sources (e.g., line lists).  Many of the eclipse depths come from the same set of observations but are analyzed using different reduction pipelines. Our study shows that subtle differences in the reduction pipelines used (e.g., the binning of WFC3 spectra into discrete channels) can drive radically different results.  This emphasizes the need for standard data sets to be used for benchmarking photometry and spectroscopy reduction pipelines.

The conflicting results of previous publications highlight the complexities of retrieval modeling and the importance of clearly communicating model assumptions. This will be especially important as retrieval models become more sophisticated with the introduction of more complicated techniques such as 3D modeling and machine learning \citep{MarquezNeilaEtal2018arxivMLRetrieval, ZingalesWaldmann2018arxivExoGAN, WaldmannGriffith2019natasMappingSaturnML, CobbEtal2019ajPlanNet}. Transparency allows for published analyses to be easily reproduced by others and encourages quicker resolution of conflicting results, which can lead to better work in the field.

Future retrieval studies should use multiple binnings of the same data to explore whether the retrievals are consistent across different binnings of the data, including the unbinned data.  This requires that data analyses publish unbinned spectra to support future retrieval studies.  Retrieval results dependent on the binning, such as those shown here, indicate the need for higher quality spectroscopic data.  A comprehensive retrieval analysis of WASP-12b with additional data from future flagship observatories, such as the James Webb Space Telescope, will provide deeper insight into the nature of this extreme exoplanet.

The Reproducible Research Compendium for this work is available for download\footnote{Available at \href{https://doi.org/10.5281/zenodo.5777204}{https://doi.org/10.5281/zenodo.5777204}.\\Note: the compendium is ~16.5 GB compressed and ~34 GB uncompressed.}.

\acknowledgments

We thank Michael Line and Julianne Moses for helpful discussions during the preparation of this manuscript. 
We also thank the anonymous referee for valuable comments that improved the quality of this manuscript.
We thank contributors to SciPy, Matplotlib, and the Python programming
language, the free and open-source community, and the NASA
Astrophysics Data System for software and services.  Part of this work
is based on observations made with the Spitzer Space Telescope,
which is operated by the Jet Propulsion Laboratory, California
Institute of Technology under a contract with NASA.  
This work was supported by NASA Planetary Atmospheres grant NNX12AI69G, NASA Astrophysics Data Analysis Program grant NNX13AF38G, and NASA Fellowship Activity under NASA Grant 80NSSC20K0682.

\bibliography{W12}

\appendix
\counterwithin{figure}{section}

\section{Figures for Electronic Supplement}
\label{sec:appwaspfigs}

The following figures are from the electronic supplement.  They are the complete set of figures for all models described in the paper.

\begin{figure*}[ht]
\includegraphics{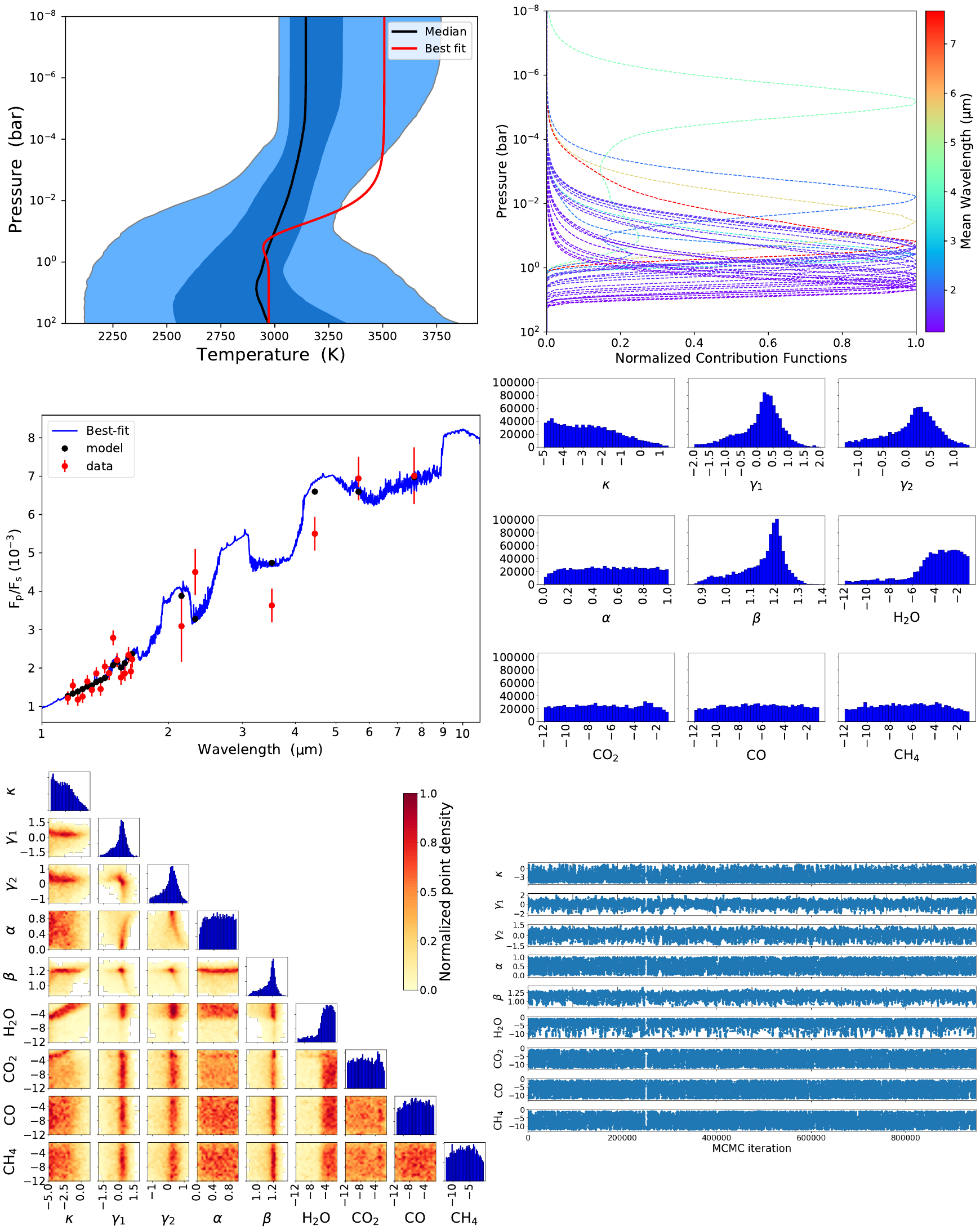}
\caption{BART results for Model 1.  \textbf{Top left:} \math{T(p)} profiles explored by the MCMC. Red line denotes the best-fit \math{T(p)} profile, the black line denotes the median \math{T(p)} profile, and the dark and light blue regions indicate the 1\math{\sigma} and 2\math{\sigma} regions, respectively. \textbf{Top right:} normalized contribution functions. \textbf{Middle left:} best-fit spectrum.  \textbf{Middle right:} 1D marginalized posteriors. \textbf{Bottom left:} 2D marginalized pairwise posteriors. \textbf{Bottom right:} trace plot for each parameter's explored values.
\label{fig:elec-supp-1}}
\end{figure*}

\begin{figure*}[ht]
\includegraphics{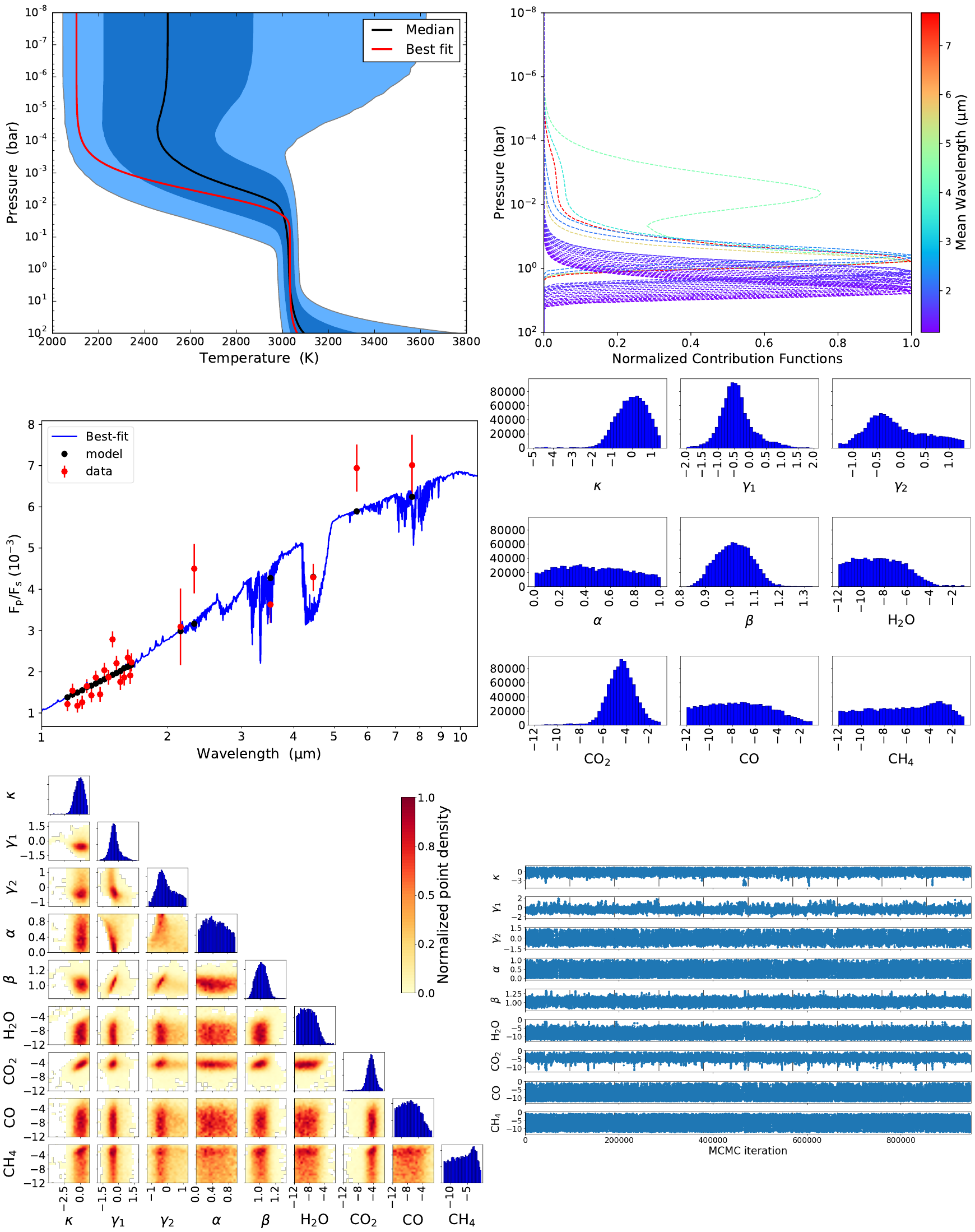}
\caption{Same as Figure \ref{fig:elec-supp-1}, except for Model 2.
\label{fig:elec-supp-2}}
\end{figure*}

\begin{figure*}[ht]
\includegraphics{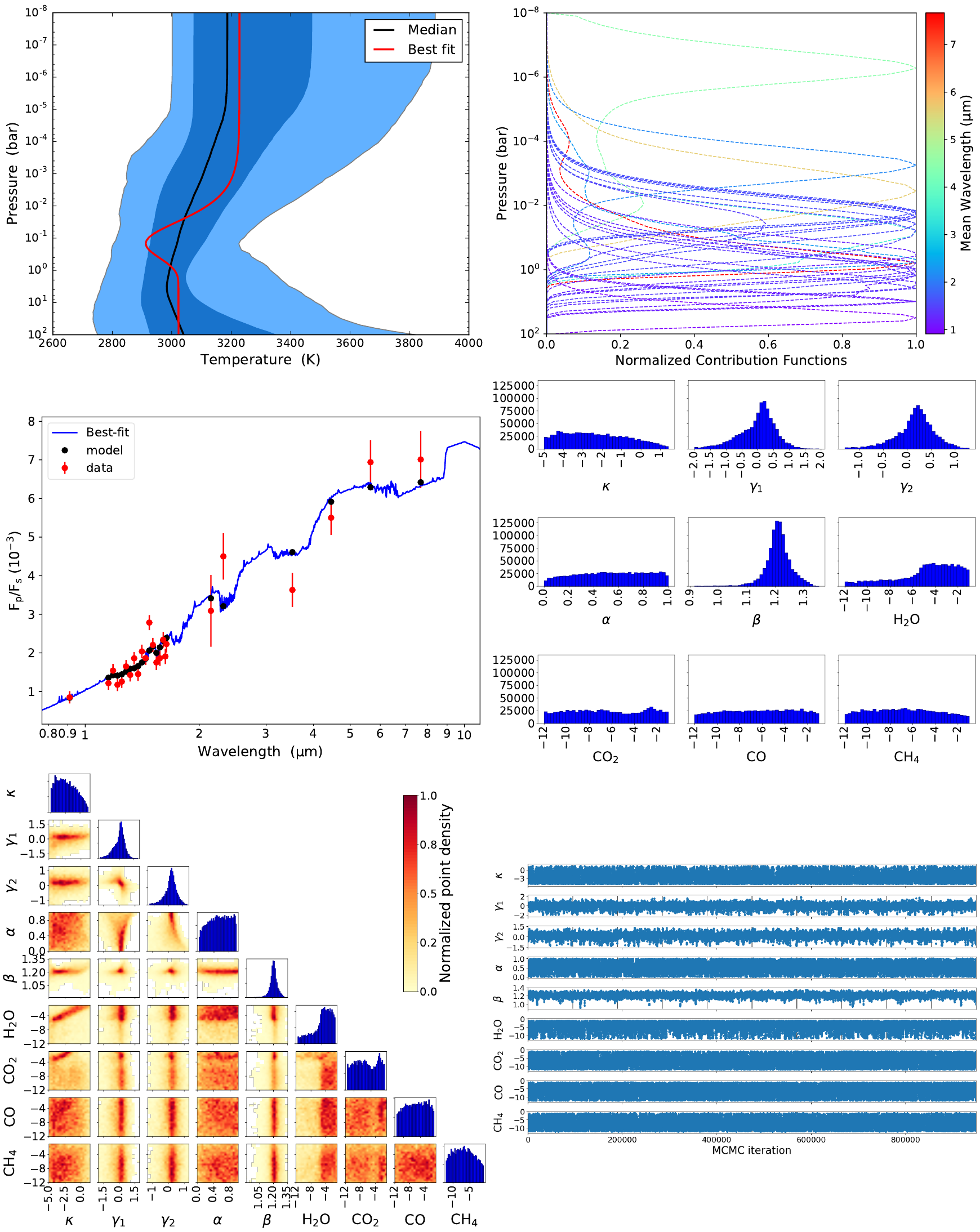}
\caption{Same as Figure \ref{fig:elec-supp-1}, except for Model 3.
\label{fig:elec-supp-3}}
\end{figure*}

\begin{figure*}[ht]
\includegraphics{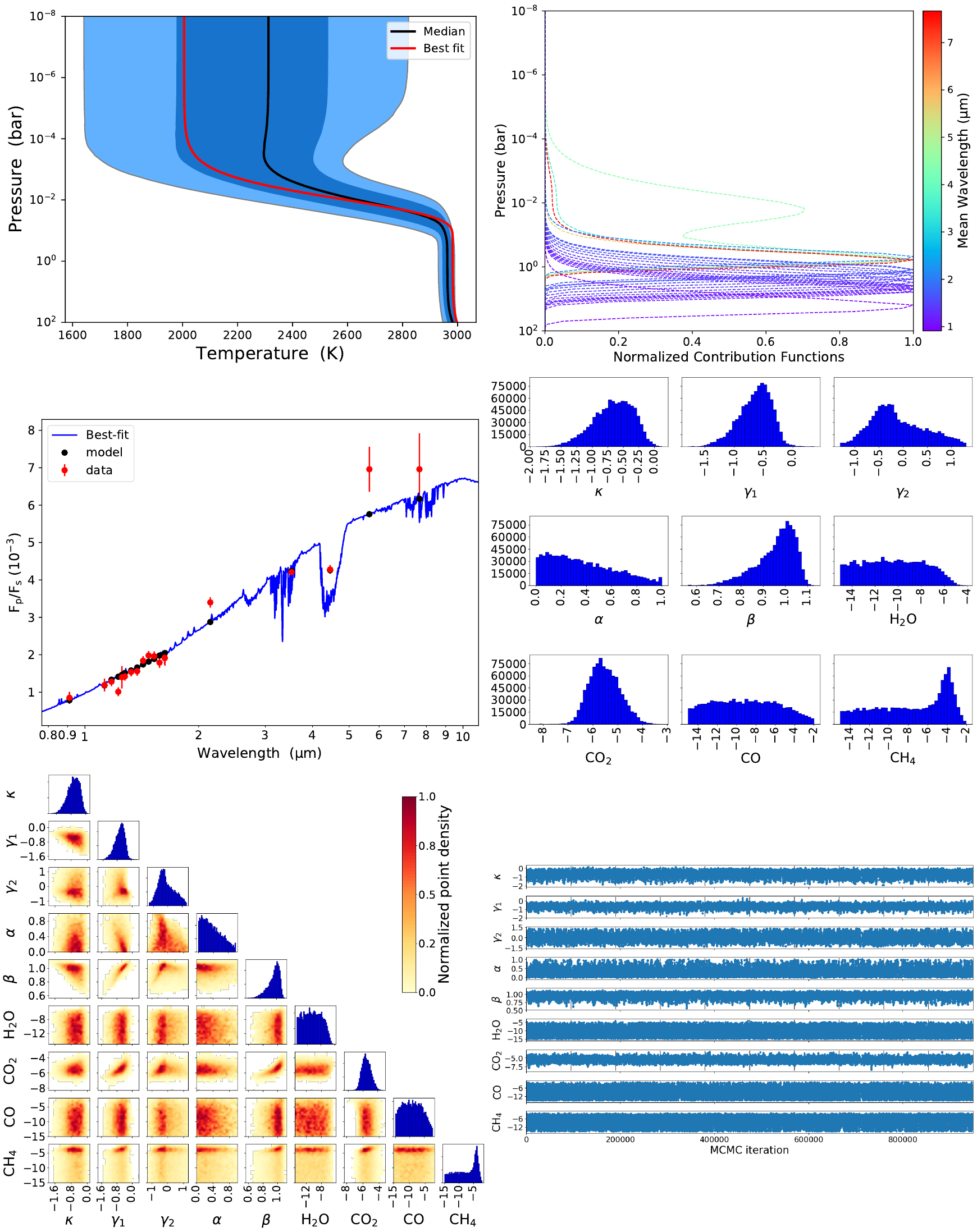}
\caption{Same as Figure \ref{fig:elec-supp-1}, except for Model 4.
\label{fig:elec-supp-4}}
\end{figure*}

\begin{figure*}[ht]
\includegraphics{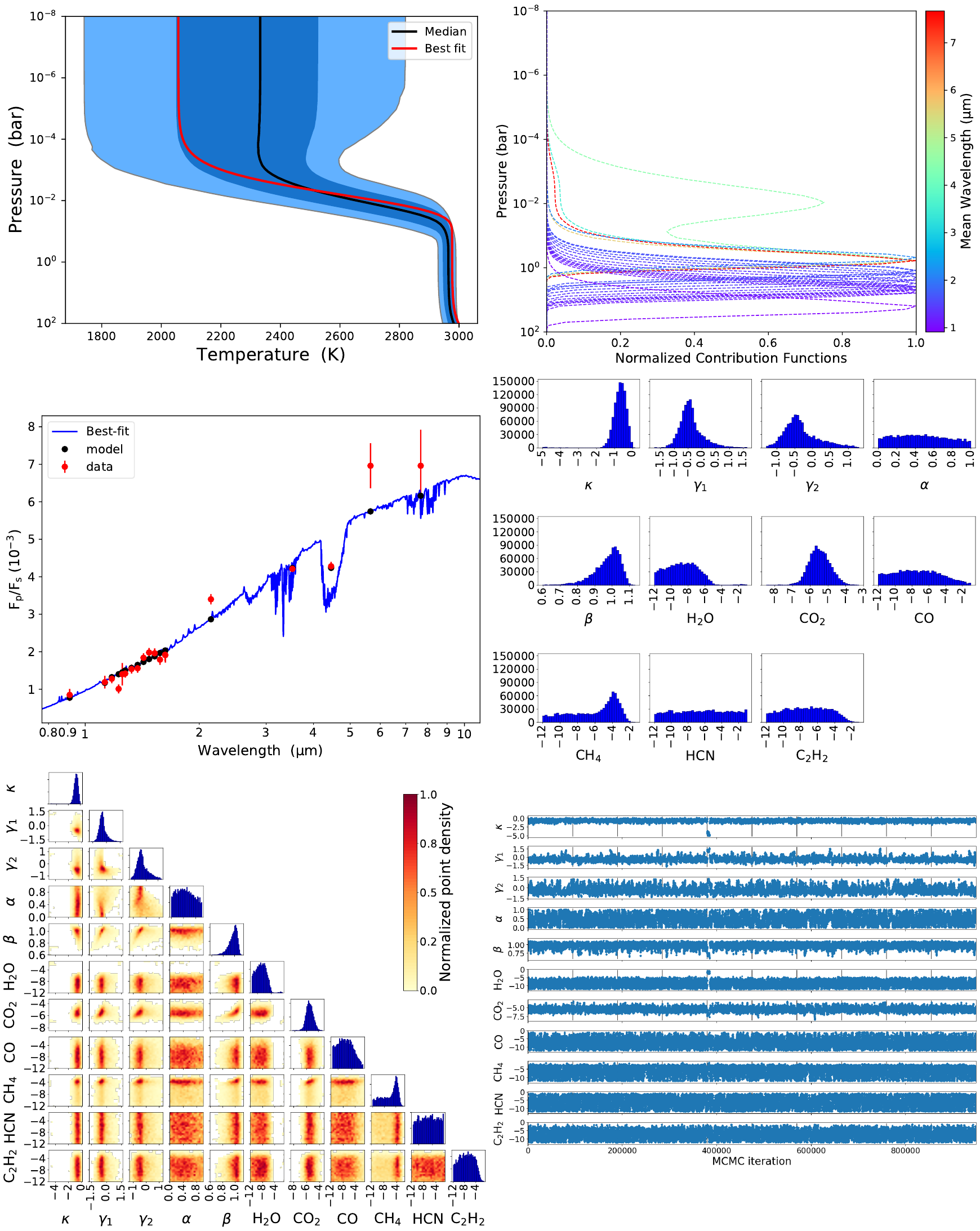}
\caption{Same as Figure \ref{fig:elec-supp-1}, except for Model 5.
\label{fig:elec-supp-5}}
\end{figure*}

\begin{figure*}[ht]
\includegraphics{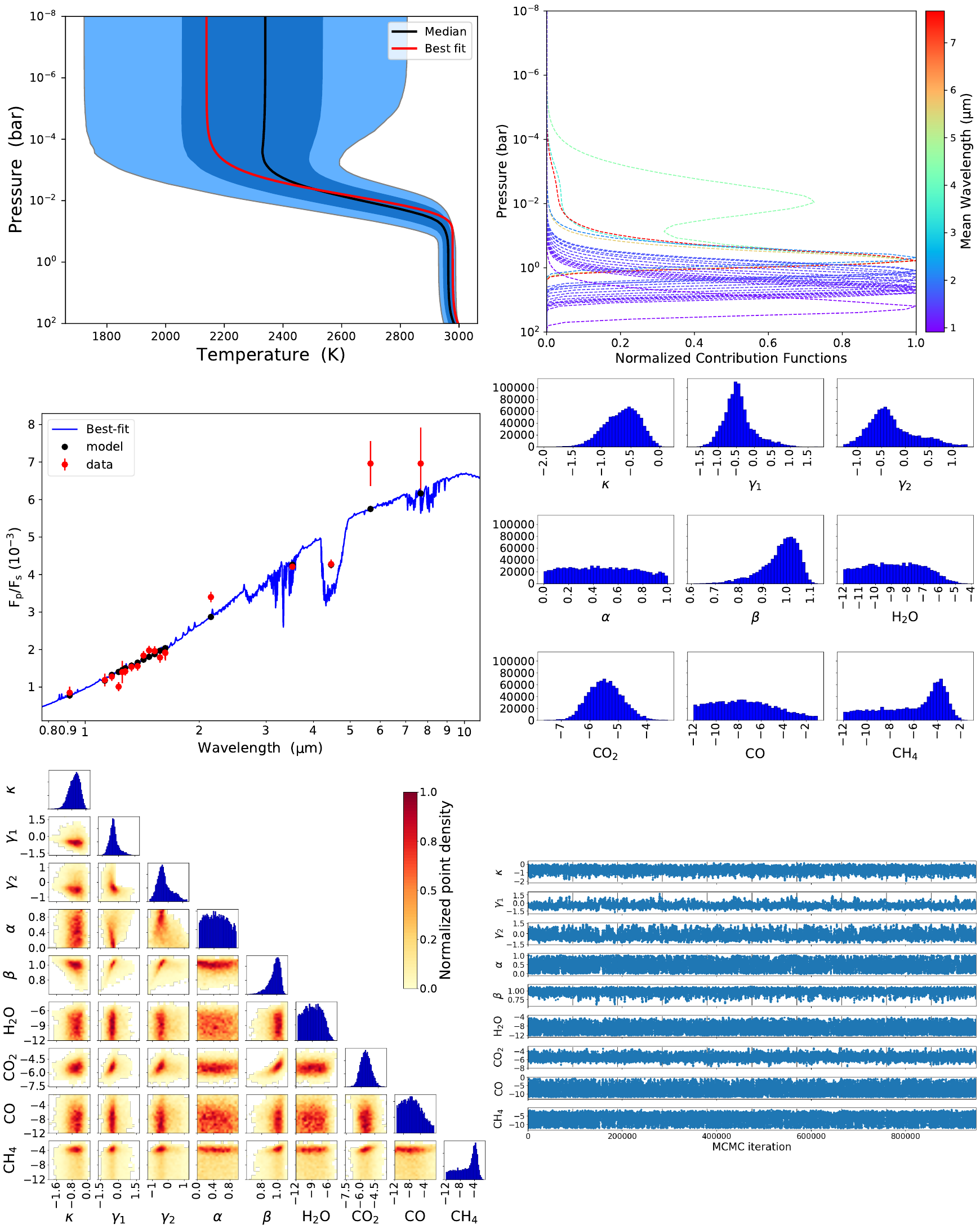}
\caption{Same as Figure \ref{fig:elec-supp-1}, except for Model 6.
\label{fig:elec-supp-6}}
\end{figure*}

\begin{figure*}[ht]
\includegraphics{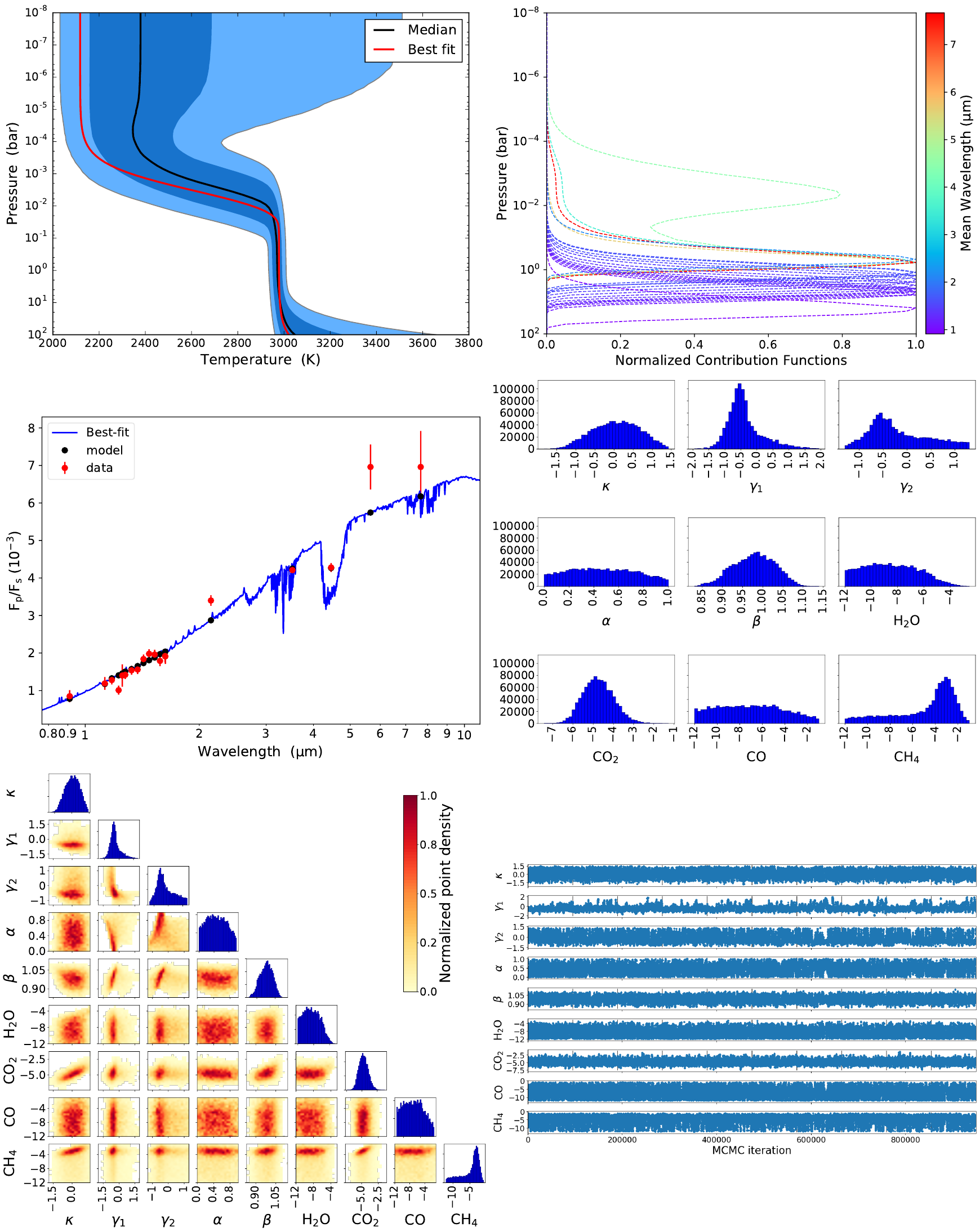}
\caption{Same as Figure \ref{fig:elec-supp-1}, except for Model 7.
\label{fig:elec-supp-7}}
\end{figure*}

\begin{figure*}[ht]
\includegraphics{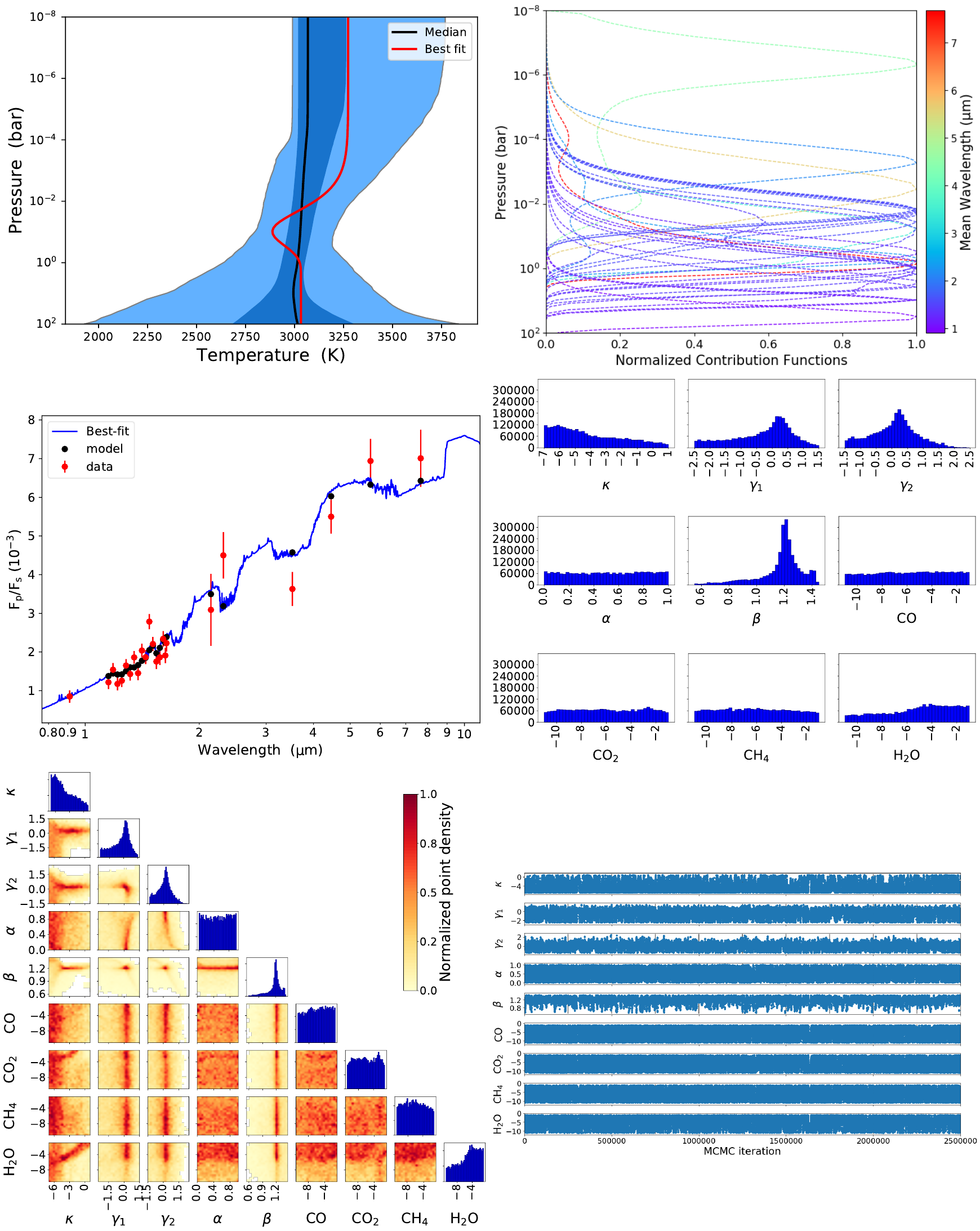}
\caption{Same as Figure \ref{fig:elec-supp-1}, except for Model 8.
\label{fig:elec-supp-8}}
\end{figure*}

\begin{figure*}[ht]
\includegraphics{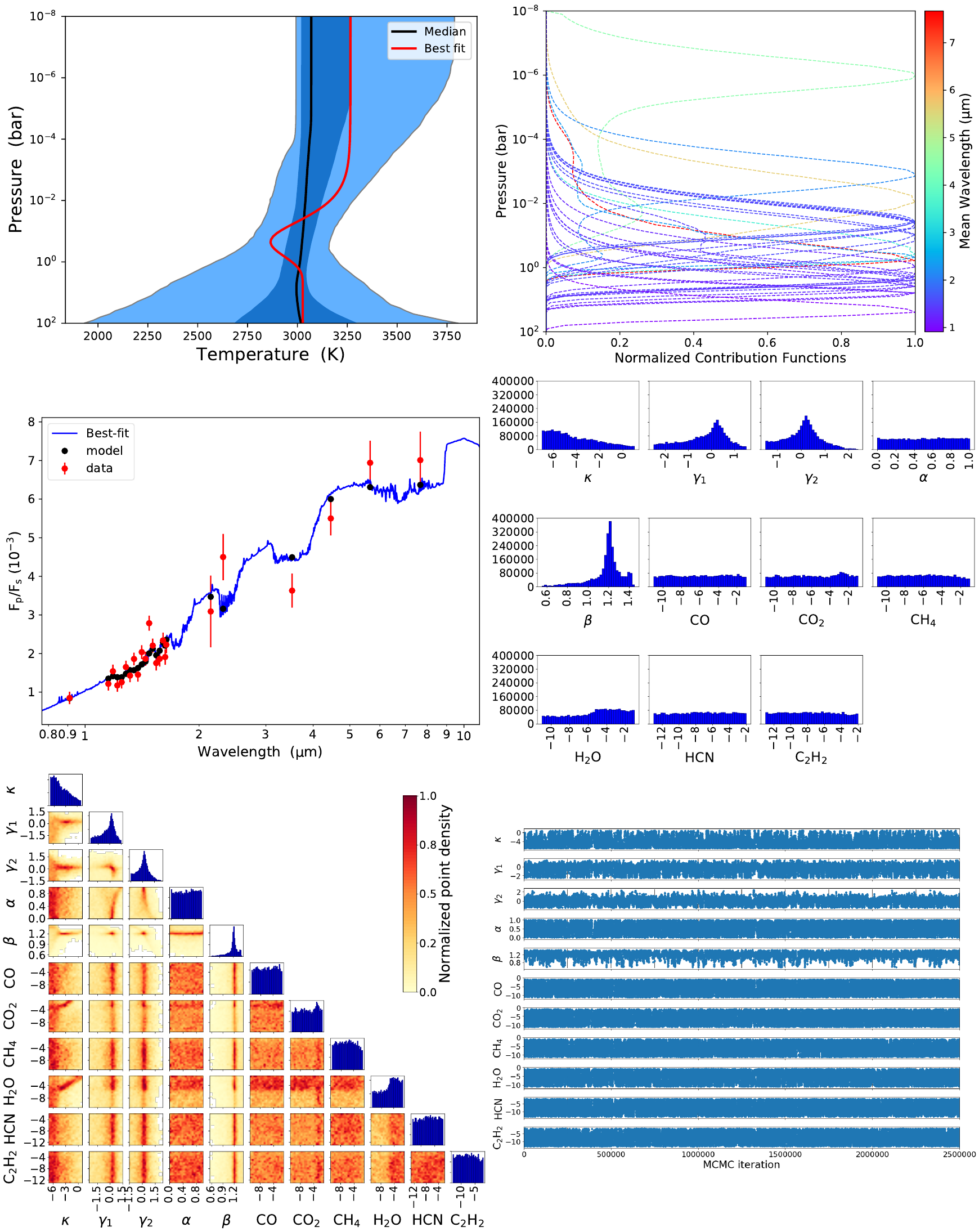}
\caption{Same as Figure \ref{fig:elec-supp-1}, except for Model 9.
\label{fig:elec-supp-9}}
\end{figure*}

\begin{figure*}[ht]
\includegraphics{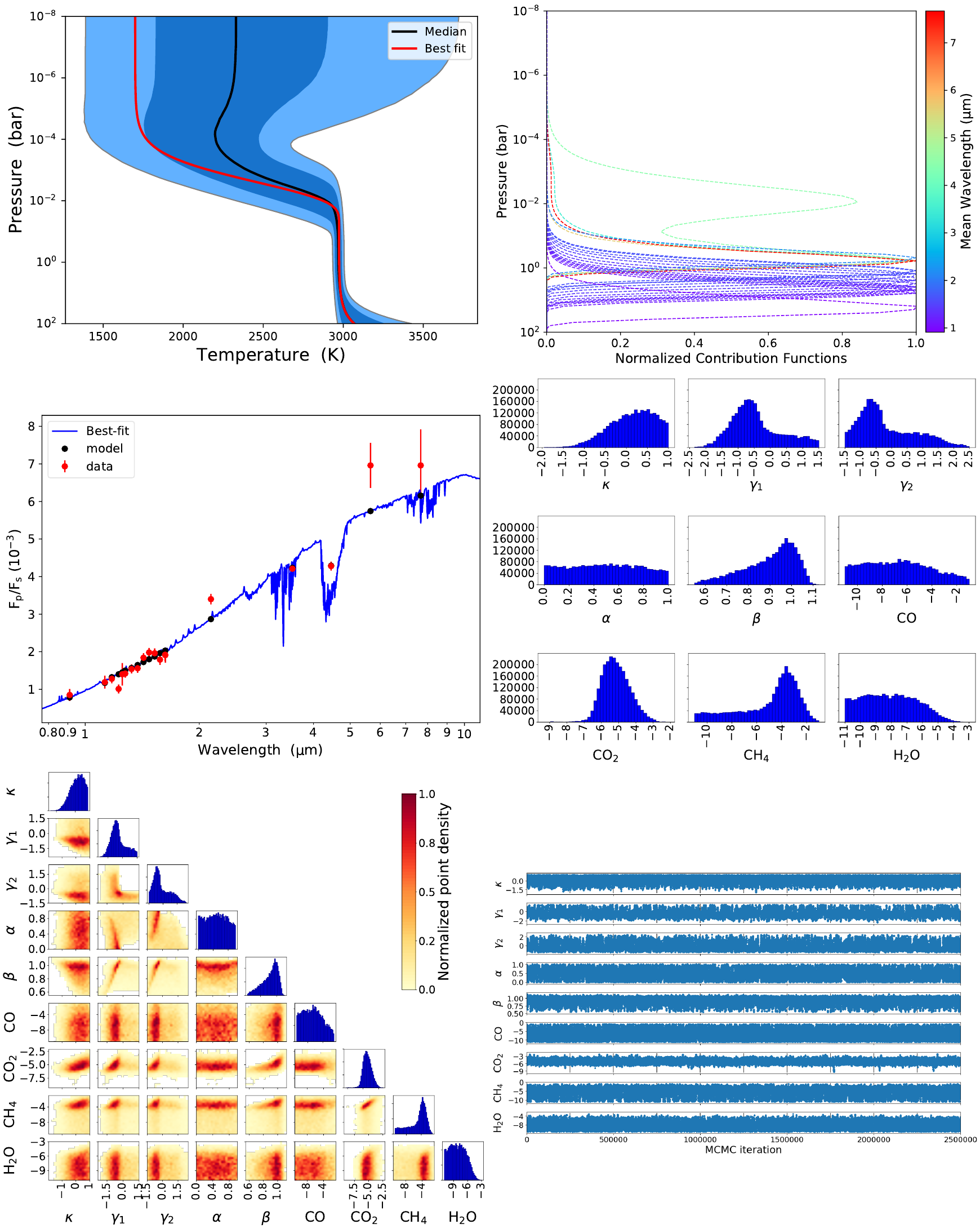}
\caption{Same as Figure \ref{fig:elec-supp-1}, except for Model 10.
\label{fig:elec-supp-10}}
\end{figure*}

\begin{figure*}[ht]
\includegraphics{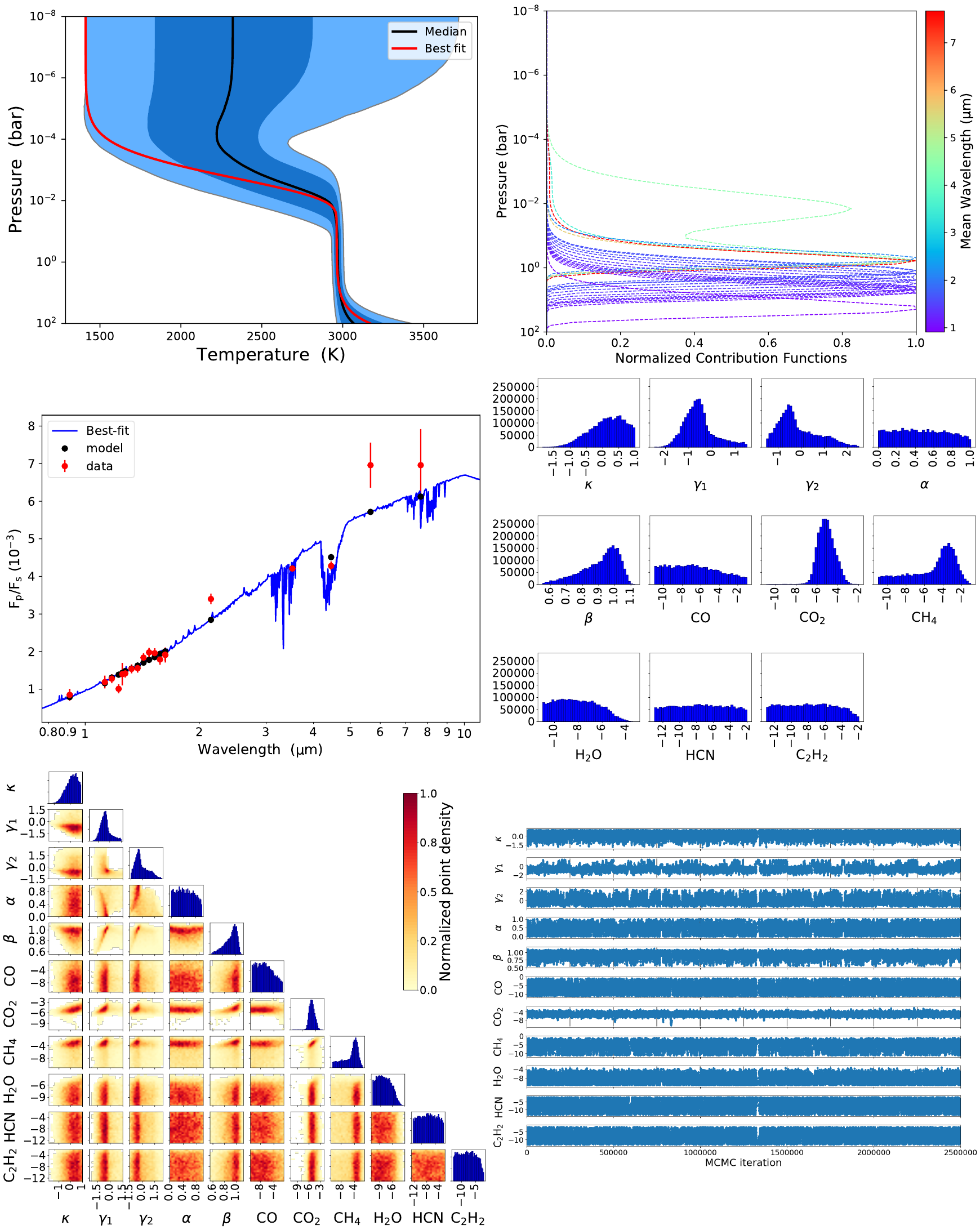}
\caption{Same as Figure \ref{fig:elec-supp-1}, except for Model 11.
\label{fig:elec-supp-11}}
\end{figure*}

\begin{figure*}[ht]
\includegraphics{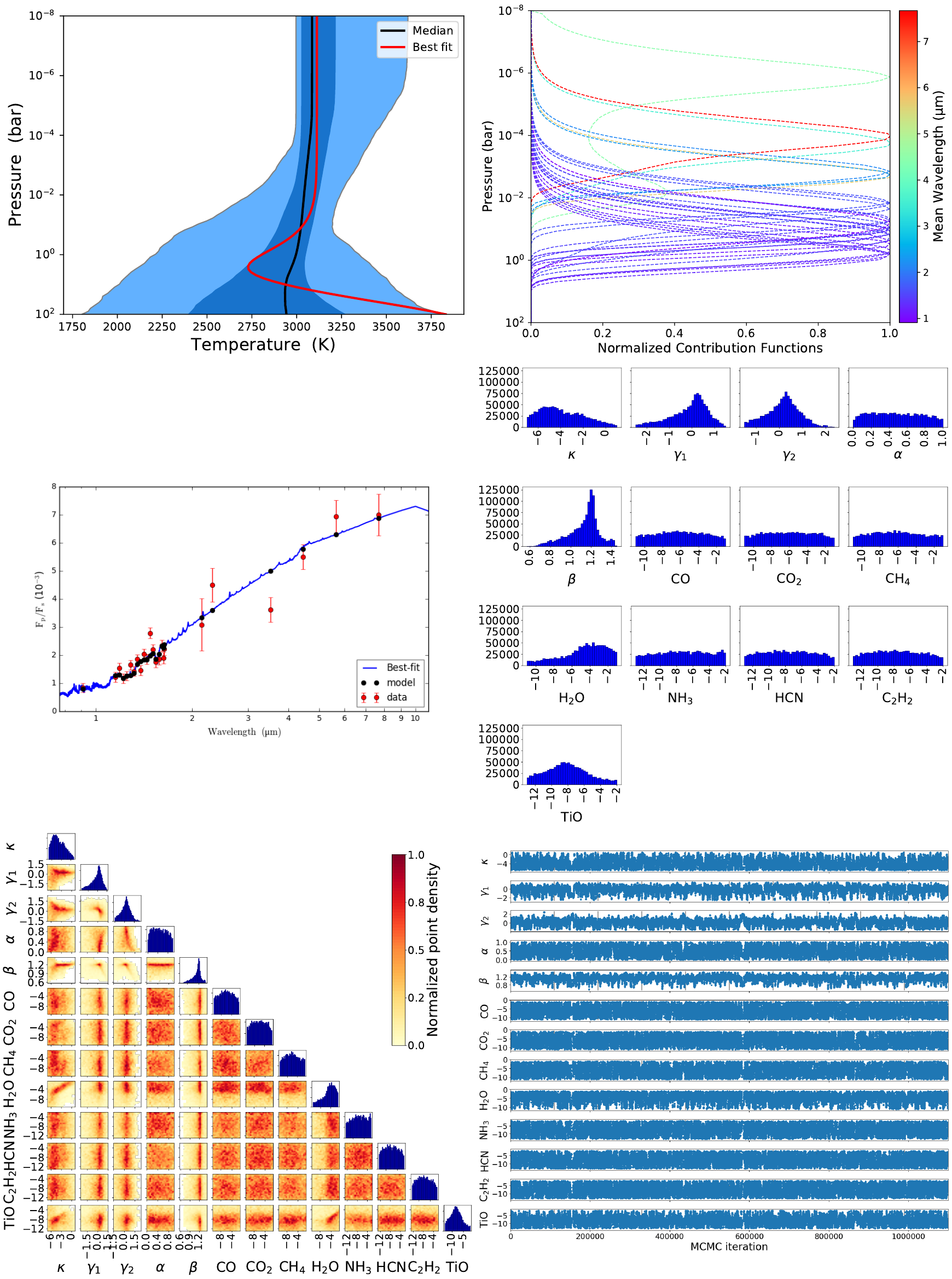}
\caption{Same as Figure \ref{fig:elec-supp-1}, except for Model \textbf{12}.
\label{fig:elec-supp-12}}
\end{figure*}

\begin{figure*}[ht]
\includegraphics{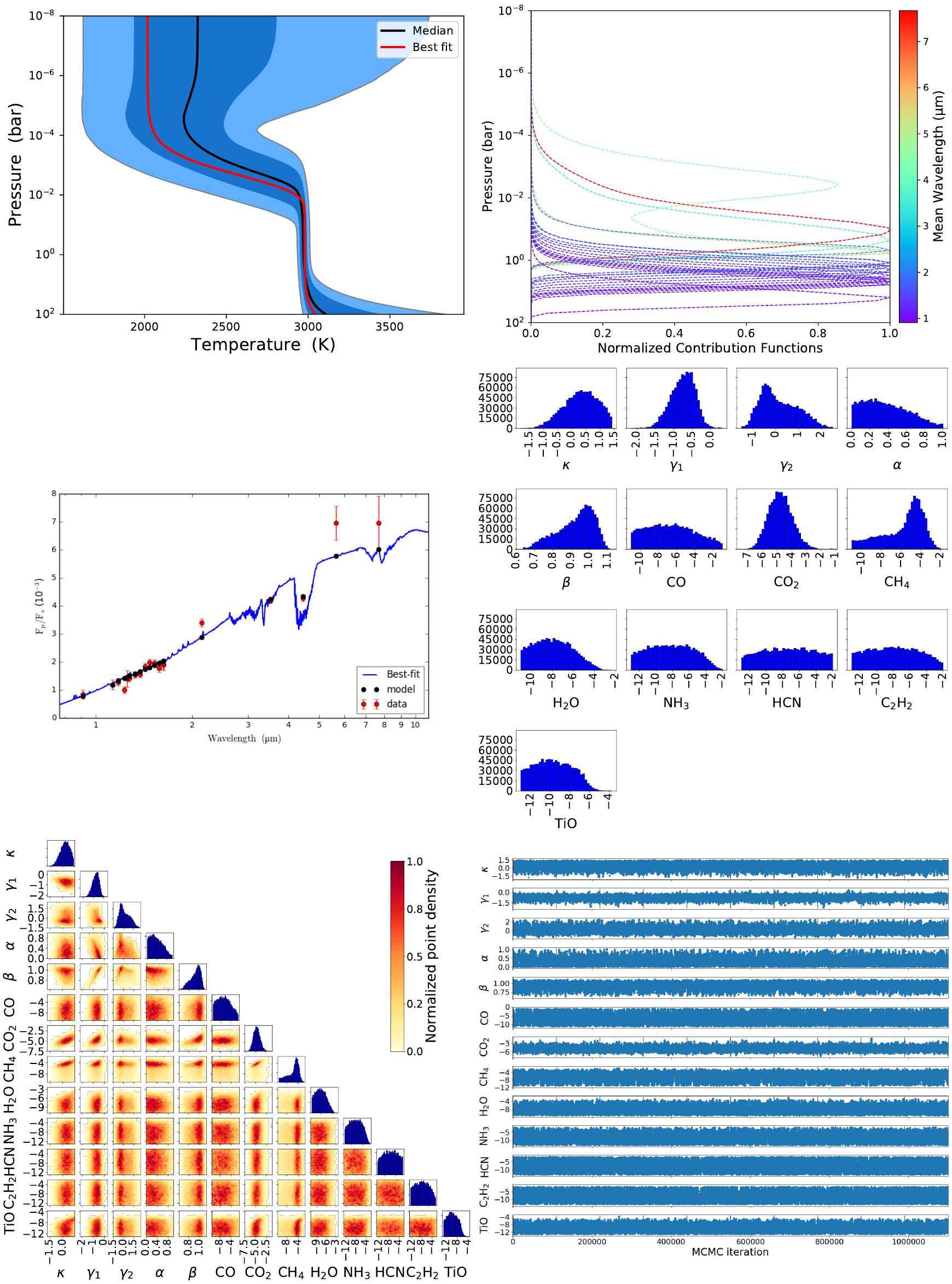}
\caption{Same as Figure \ref{fig:elec-supp-1}, except for Model \textbf{13}.
\label{fig:elec-supp-13}}
\end{figure*}

\end{document}